\documentclass[preprint,pteplogo]{ptephy_v2}

\preprintnumber{XXXX-XXXX} 
\usepackage{hyperref}
\usepackage{longtable}
\usepackage{siunitx}
\usepackage{xcolor}

\begin{document}

\title{Two-stage Convolutional Neural Network for pseudo six-dimensional phase space reconstruction}




\author{S.~Mukherjee}
\author{M.~Kuriki}
\author{Z.~J.~Liptak}
\affil{Graduate School of Advanced Sciences of Matter, Hiroshima University,\\
       1-3-1 Kagamiyama, Higashi-Hiroshima, Hiroshima 739-8530, Japan\\
       \email{sayantan.96.phy@gmail.com, d223836@hiroshima-u.ac.jp}}

\author{H.~Hayano}
\affil{NovAccel Co.\ Ltd., 3706-1 Mushikake, Tsuchiura, Ibaraki 300-0066, Japan}

\author{M.~Kurata}
\author{N.~Terunuma}
\author{T.~Okugi}
\author{Y.~Yamamoto}
\affil{Accelerator Laboratory, High Energy Accelerator Research Organization (KEK)\\
       1-1 Oho, Tsukuba, Ibaraki 305-0801, Japan}




\begin{abstract}
  
In particle accelerators, broad characterization of the six-dimensional (6D) beam phase space is crucial but difficult to obtain with conventional beam diagnostics. We develop a two-stage convolutional neural network (CNN) that reconstructs the 6D phase space from only sixteen transverse $x-y$ screen images taken at a place with dispersion by different phase space rotation angles. The model is trained with simulation data of KEK-Accelerator Test Facility (ATF) injector with ASTRA. The real-space images in the chicane orbit at the KEK-ATF injector were acquired by varying the RF phase of the RF electron gun and the solenoid magnetic field. From these data, we reconstructed a pseudo 6D phase space distribution at the cathode surface, expressed through 15 two-dimensional (2D) distributions covering all pairwise coordinate combinations. The time width and spatial spread of the electron beam at the cathode showed values consistent with the measured values at KEK-ATF. Compared to existing 6D beam imaging measurement techniques such as tomography, it significantly reduces measurement time and required computational resources, enabling the provision of a more practical 6D phase space measurement method.

\end{abstract}

\subjectindex{xxxx, xxx}

\maketitle


\section{Introduction}

Beam quality, usually quantified by its transverse and longitudinal emittances, plays a key role in accelerators. In linear colliders, ultra-low transverse emittance at the interaction point is required to form extremely flat, nanometer-scale beams and thereby achieve the design luminosity~\cite{Behnke2013}.
Synchrotron light sources benefit from emittances pushed below the diffraction limit, which increases the spatial coherence and significantly enhances the spectral brightness of the photon beam~\cite{Wiedemann2003}. For free electron lasers (FEL), larger emittance leads to increased beam divergence and effective energy spread in the undulator, degrading the FEL gain~\cite{Huang2006}. In medical accelerator facilities, such as cyclotrons used for proton therapy to treat tumors, a well-controlled beam shape is crucial for transporting and focusing $70-250\,\mathrm{MeV}$ proton beams~\cite{Maradia2025}. 

Meeting the increasingly demanding operating conditions of modern accelerator facilities, such as those listed above, requires the ability to assess and adjust optics for run-time beam conditions. 
Reconstructing the beam phase space along the beamline is therefore highly valuable for diagnosing beam quality.
In general, relying only on emittance is insufficient, since it does not capture complex features of the 6D phase space distribution like multiple peaks, correlation among the degrees of freedom, etc.  
Access to the 6D phase space distribution enables a more complete evaluation of beam performance and provides insight into the mechanisms responsible for beam quality degradation. However, conventional reconstruction approaches within an accelerator are often difficult and time-consuming, requiring destructive measurements, computationally intensive back-projection, and typically providing access to only a limited number of phase space dimensions.

Numerous conventional beam-diagnostic techniques have been developed to infer transverse phase space.These include pepper pot or multi slit emittance monitors~\cite{Wang1990,Wang1991}, nano fabricated wire scanners~\cite{Hermann2021}, and laser wire scanners~\cite{Jcwang2022}, among others. In standard implementations, such diagnostics can reconstruct the projected 2D transverse phase spaces $(x,x')$ and $(y,y')$, and under well-controlled conditions, they can also infer a coupled 4D transverse beam matrix. However, they primarily probe transverse dynamics and do not directly provide access to the 6D phase space distribution. As a result, correlations between transverse and longitudinal degrees of freedom and other higher-dimensional structures are not uniquely captured.

In tomography techniques~\cite{mckee1995,hancock1999,stratakis2007,stratakis2008,yakimenko2003,hock2013,scheins2004,townsend1993,wong2022}, the phase space distribution is inferred after varying the beam optics. In this case, projections are acquired at different rotation angles, and the underlying distribution can be reconstructed using back-projection methods. However, this approach requires precise magnet settings corresponding to a large number of projection angles, making implementation difficult on machines without a dedicated tomography setup. The analysis of up to four dimensions of phase space can be achieved by these methods.

Although the conventional techniques or tomographic reconstructions served as useful tools in phase space evaluation, as the complexity becomes much greater in higher dimensions, only a few studies have been done regarding the 6D phase space reconstruction. One experimental measurement was performed at the Spallation Neutron Source (SNS) Beam Test Facility (BTF) in 2018~\cite{Cathey2018}. A dedicated experimental setup was used that contained six movable slits to gather the information of particles within a fractional region of the 6D phase space. The measurement was highly time-consuming and required 32 hours of constant beam current. The result of the study showed the existence of correlations among the transverse and longitudinal degrees of freedom in the 6D phase space, which are driven by the Coulomb force.

Recently, machine learning has gained importance in various sectors like image classification, speech recognition, data mining, etc~\cite{wang2019,ankerst1999,fujiyoshi2019,he2016,baldi2011}. Some studies are performed using these neural network algorithms for phase space reconstructions~\cite{roussel2024,wolski2022,scheinker2023,emma2018}. Among these, one very recent study~\cite{roussel2024} showed the reconstruction of 6D phase space using the generative phase space reconstruction (GPSR) technique, which uses backward-differentiable simulations and optimizes neural network parameters to generate the 6D distribution. In this framework, backward differentiability allows the initially guessed 6D beam distribution to be iteratively refined so that it agrees with the experimental measurements. It employs a loss function based on the negative logarithm of the 6D beam emittance, which is proportional to the entropy of the distribution. Following the principle of maximum-entropy tomography (MENT)~\cite{KMHock2013}, the likelihood of the reconstructed distribution is maximized when its entropy is maximized. It gave consistent results and could reconstruct many complex planes of the 6D phase space. But the biggest limitation of the study is the fact that it fully relies on backward-differentiable particle tracking, which is not available in standard simulation packages, and one has to modify existing software. Also, it needs the available experimental images to compute many solutions for maximizing the entropy and high computation from an A100 NVIDIA GPU in order to get to a 6D distribution.

Our study is based on developing an AI model for a typical injector beamline using a Convolutional Neural Network (CNN)~\cite{alzubaidi2021,gu2018,yao2019,simonyan2014} to solve the inverse problem of generating a pseudo 6D phase space distribution from measured 2D real space beam images. The evaluated distribution is pseudo due to the fact that the reconstruction is based on 15 pairwise 2D phase space distributions, which do not uniquely determine the full 6D phase space density because higher order correlations among three or more phase space coordinates remain unresolved. Using successive convolution operations, the CNN acts as a feature extractor for the input beam images, capturing information such as the beam size, overall shape, and how the intensity is distributed across the image. The method is tomographic in the practical sense that it uses a limited set of machine scans, here implemented by varying the RF gun phase and solenoid field to provide multiple informative views. However, it does not require dense $0^\circ$--$360^\circ$ angular coverage or a large number of optics settings as in conventional back-projection tomography. Moreover, unlike GPSR-type approaches, the CNN method can be trained using readily available forward simulation codes to generate the required datasets. Although the training takes a fairly large dataset, using about $5\times10^4$ images, it can be carried out on a modest GPU. Once the model is fully trained, the reconstruction takes less than a minute to finish. Such a shorter reconstruction time makes the model highly desirable for accelerator facilities, and it can be used as an online beam diagnostic tool during the experimental beam time. To overcome the common problem of a finite training set for machine learning models, we utilize Fourier series functions that cover most of the possible beam distributions. By providing a wide variety of beam shapes in training, we improve the model's ability to extrapolate beyond the training set. Furthermore, if needed, it can be fed with higher-order Fourier series distributions to improve the quality. We demonstrate that, given a wide variety of beam shapes in training, the model can learn the beam complexity to reconstruct the 6D phase space.

In this paper, we present a newly developed CNN-based AI model and evaluate its performance on both synthetic beams and KEK-ATF experimental data. The principles of phase-space rotation necessary for tomography are described in Section~\ref{sec:principles}, and a detailed description of the CNN model is given in Sec.~\ref{sec:model_description}, where we describe the network architecture and the training process. Section~\ref{sec:synthetic_samples} shows the performance on the synthetic beam distributions. An experimental demonstration of the technique at the KEK-ATF injector beamline is presented in Sec.~\ref{sec:experiment}.
Finally, Sec.~\ref{sec:discussion} and Sec.~\ref{sec:conclusion} summarize the study and discuss the prospects for the accelerator community.

\section{Principles of phase-space rotation and measurement for beam reconstruction}
\label{sec:principles}

Reconstruction of the 6D phase space of the beam can be achieved by observing changes induced by phase-space rotations of the beam in flight. Typically, reconstruction focuses on the transverse dimensions due to the relative difficulty of measuring changes in the longitudinal dimensions and the presence of complex nonlinear effects which can make calculation cumbersome or impossible. In this work, we demonstrate that a small number of phase-space rotations performed with common beamline optical elements and RF phase modulation can effect the necessary phase-space rotations for our CNN-based algorithm to do the 6D phase space reconstruction, provided that the beam's transverse ($x-y$) distribution can be measured in an area with observable dispersion.
Changes induced by the RF phase differences are observable in regions with dispersion, in our case chosen to be at the center of the chicane. By bending the beam in the $x$ direction, the chicane dipole magnets induce an asymmetric effect: the beam energy spread becomes visible as beam spread in the $x$ dimension, while edge focusing effects induce changes in the particle trajectories in the $y$ direction~\cite{mukherjee2023}. We take advantage of these effects to observe changes in both the transverse and longitudinal dimensions introduced by scanning the solenoid and RF phases, respectively.
Measurement is performed with a phosphor screen inserted into the chicane region, providing data on beam shape and intensity. 

To obtain the phase-space rotations necessary to demonstrate our technique, we employ changes in the solenoid field and RF phase offset with respect to the laser pulse timing, rotating the transverse and longitudinal dimensions, respectively. These rotations are briefly described below.

\begin{figure}[httb]
	\centering
	\includegraphics[width=0.45\linewidth]{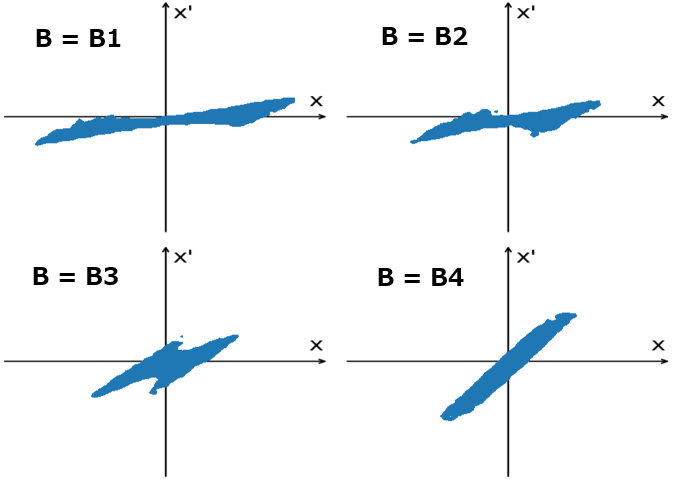}
	\caption{Rotation of $x-x'$ phase space by varying the solenoid peak field, $B$. The motion is influenced by solenoid focusing, dispersion due to chicane and space charge effects.}
	\label{fig:schematicxx'}
\end{figure}

\subsection{Transverse phase-space rotation}

\begin{figure}[httb]
  \centering
  \includegraphics[width=0.45\linewidth]{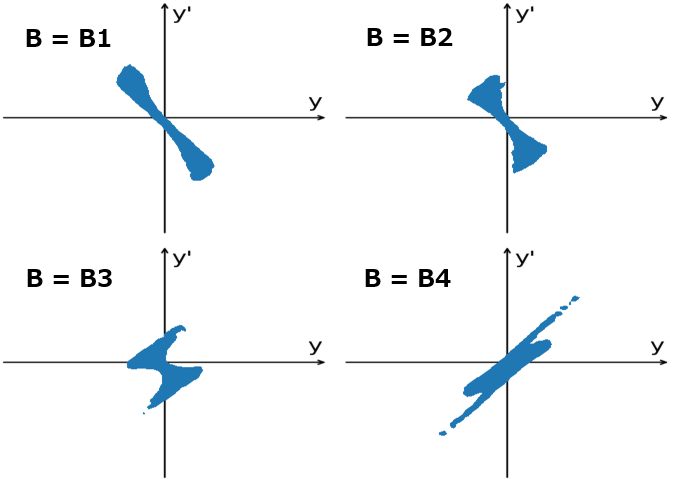}
  \caption{Rotation of $y-y'$ phase space by varying the solenoid peak field, $B$. The overall effect comes from solenoid focusing, edge focusing by chicane and space charge effects.}
  \label{fig:schematicyy'}
\end{figure}

The solenoid field effects rotations in the transverse directions, along with focusing and partial compensation of space charge effects. Because the solenoid focuses in both transverse directions simultaneously, a single field is sufficient to induce changes in both the $x-x'$ and $y-y'$ planes. Changes in the transverse phase spaces as a result of variations in the solenoid field are shown in Figs.~\ref{fig:schematicxx'} and~\ref{fig:schematicyy'}. Focusing from the solenoid alone is symmetric in the $x$ and $y$ dimensions, but an asymmetry is introduced at the measurement point due to edge focusing effects in $y$ and beam spreading due to bending in $x$. Figures~\ref{fig:schematicxx'} and~\ref{fig:schematicyy'} reflect the observed phase-space rotations within the measurement region of the chicane.

\subsection{Longitudinal phase-space rotation}

\begin{figure}[httb]
  \centering
  \includegraphics[width=0.45\linewidth]{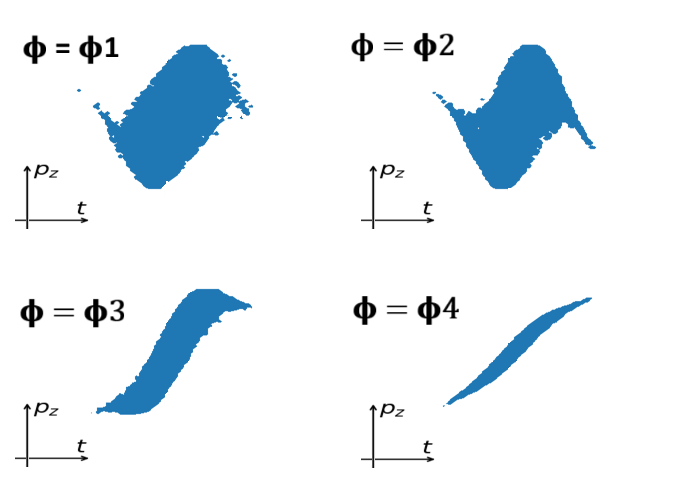}
  \caption{Rotation of $t-p_z$ phase space in the presence of different RF phases ($\phi$). The net effect is due to the change in energy spread and space charge forces.}
  \label{fig:schematictpz}
\end{figure}

Rotation in the longitudinal ($t-p_z$) phase space plane is effected by varying the RF phase offset at the cathode. As a result, beam bunches in differing RF phases are accelerated at different points in the RF wave, affecting the overall beam energy spread. Figure~\ref{fig:schematictpz} shows examples of rotations in the $t-p_z$ plane as a result of RF phase changes. 

\section{CNN Algorithm for 6D phase space reconstruction}\label{sec:model_description}
We employ a new CNN algorithm to reconstruct the 6D phase space of the beam on the cathode using the principles described in Section~\ref{sec:principles}. It is a specialized type of deep learning model designed primarily for processing and analyzing visual data~\cite{alzubaidi2021,shea2015}. CNNs preserve the spatial structure of images, allowing them to efficiently capture local patterns and visual features.
The core component of a CNN is the convolutional layer, which applies small filters, also called kernels, across the input image. These filters slide over the image and perform mathematical operations to detect features such as edges, corners, textures, and shapes. As the data passes through deeper layers, the network learns more complex and abstract representations, enabling it to recognize objects, faces, or scenes. The algorithm also consists of Pooling layers that are used to reduce the spatial size of the images, which helps decrease computational cost and improves robustness to small shifts or distortions in the input. Finally, fully connected layers interpret the extracted features and produce classification or prediction results. Our algorithm processes beam images using a CNN, and instead of extracting features, it generates multiple images. From the real-space images obtained by beam measurement, we reconstruct two-dimensional images of any two variables in six-dimensional phase space.

We train the network using conventional forward ASTRA~\cite{flottmann2017} simulations, with no requirement for backward-differentiability or code modification. The model also does not solve a maximum-entropy optimization problem at inference time; once trained, it maps measured chicane $x-y$ images to the cathode 6D phase space in well under a minute on a comparatively modest, affordable GPU. The convolutional architecture extracts image features at the chicane measurement point and learns their nonlinear relationship to the upstream 6D phase-space coordinates at the photocathode, providing a more practical route to high-dimensional beam reconstruction.
 
The previous studies done on phase space tomography~\cite{mckee1995,hancock1999,stratakis2007,stratakis2008,yakimenko2003,hock2013,scheins2004,townsend1993,wong2022} involved back-projecting the lower-dimensional projections to the phase space distribution. However, in such techniques, the reconstruction highly depends on the number of projections that can cover the angular range from $0^\circ$ to $360^\circ$ to describe the phase space distribution. Furthermore, these methods, based on the principle of Fourier slice theorem~\cite{Fangyi2010} and MENT, use one-dimensional projections, which also reduces higher-dimensional information of the phase space. In contrast, our CNN model uses the full 2D beam images as projections, preserving richer phase space features and not requiring coverage over the full angular range.

As mentioned in the Introduction, the recent GPSR study~\cite{roussel2024} demonstrates that machine-learning-based generative models can, in principle, reconstruct the 6D phase space from 2D beam images without strict constraints on projection angles. However, this approach has two practical drawbacks. First, it relies on fully backward-differentiable particle-tracking, which is not available in standard accelerator simulation packages and requires substantial modification of existing codes. Second, the optimization is computationally intensive: maximizing the MENT-based objective over a 6D distribution demands long runs on a dedicated A100 NVIDIA GPU, hardware that is expensive and not easily accessible.

Our CNN-based approach overcomes these limitations. Below, we describe the algorithm's details, its training, the reconstruction of test data from simulations, and the reconstruction of experimental data at KEK-ATF. 

\subsection{Model Architecture}

\begin{figure*}[httb]
  \centering
  \includegraphics[width=\textwidth]{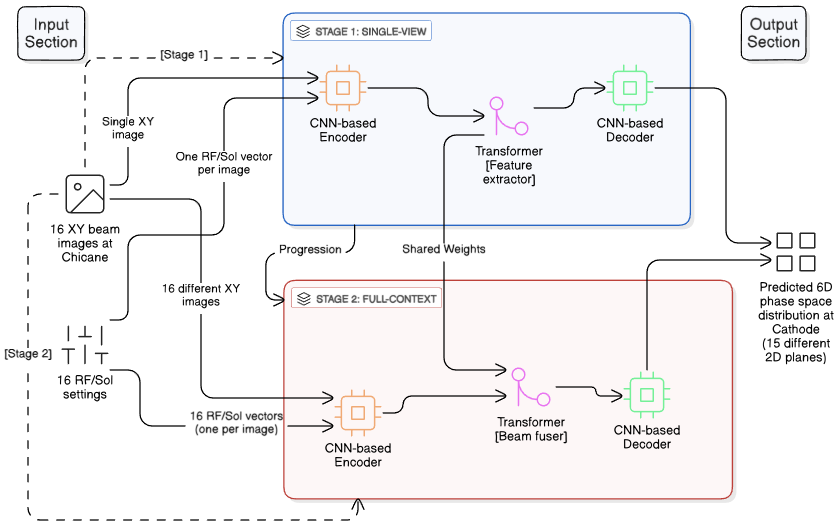}
  \caption{The detailed network architecture of the two-stage CNN model. In Stage~1, the model learns from one beam image at a time while keeping the RF phase and solenoid setting fixed, and this training is repeated across all cathode phase spaces and multiple RF-solenoid configurations. In Stage~2, the model takes all 16 chicane beam images together with their RF-solenoid settings and learns to combine them to predict a single cathode phase space distribution.}
  \label{fig:cnn_2stage}
\end{figure*}

Our network has three main parts: an encoder, a transformer, and a decoder, trained in two stages as shown in Fig.~\ref{fig:cnn_2stage}. For a given RF phase and solenoid field, the encoder~\cite{michelucci2022} takes a single-channel $64\times64$ histogram of the $x-y$ image at the chicane measurement point and passes it through three convolutional layers with pooling~\cite{shea2015}, gradually compressing the image into a $4\times4$ feature map with 128 channels. This feature map can be viewed as a small grid where each cell contains 128 learned numbers that summarize local patterns in the image, such as beam size, position, and shape. Together, this feature map is a compact representation of the original $64\times64$ image. The feature map is then flattened and passed through two fully connected layers, which mix these features and compress them into a 150-dimensional image latent summarizing the beam image.

In parallel, the RF phase and solenoid field, the knobs used to generate the chicane image, are treated as a 2-dimensional input vector and passed through four fully connected layers to produce a 150-dimensional knob latent. This knob latent is concatenated with the image latent and processed by two fully connected layers with a ReLU~\cite{agarap2018} nonlinearity in between, producing a single 150-dimensional “projection embedding”. This embedding is a joint representation of both the RF-solenoid setting and the measured $x-y$ image at the chicane.

For each sample in the dataset, we have (i) the distribution at the cathode that defines 15 two-dimensional projections of the 6D phase space, (ii) 16 different RF-solenoid pairs, and (iii) 16 corresponding real-space beam images at the chicane. Using the encoder-knob fusion described above, each of the 16 chicane images is mapped to a 150-dimensional projection embedding. These 16 embeddings are then fed into the transformer, which applies three self-attention layers with six heads. Each attention head learns how to weight and combine information across all 16 views, and the final transformer output is collapsed into a single “beam representation” vector of dimension 1280.

In Stage~1, we train the model using individual chicane $x-y$ images at a time together with a fixed RF and solenoid field. 
The encoder combines each image and the RF-solenoid settings into a feature vector, and the decoder predicts the 15 cathode phase-space histograms from it using the transformer~\cite{vaswani2023}.
The goal of this stage is to teach the network the basic inverse mapping: how changes in the cathode distribution appear in the chicane image when the beamline configuration is fixed.
For consistency with Stage~2, we feed the transformer a sequence of length 16 by replicating the same single-view feature (with positional encodings), but the input still represents only one measured view.
This single-view pretraining provides a stable initialization before Stage~2 learns to combine information from all 16 views.

In Stage~2, we switch to the full multi-view setting. 
For each sample, the transformer receives all 16 chicane $x-y$ images together with their corresponding RF phase and solenoid field values, and it is trained to reconstruct one consistent cathode 6D phase-space distribution (15 output histograms).
The key idea is that the same cathode distribution produces different chicane images when the RF-solenoid settings are changed.
By learning how these 16 views relate to the same underlying beam, the network can combine their complementary constraints and reduce the ambiguity of single-image inversion, enabling a more reliable recovery of the cathode distribution.

After the transformer, the decoder~\cite{michelucci2022} maps the 1280-dimensional beam representation back into physical space. It first expands this vector into a coarse feature map and then applies three upsampling convolutional layers to reconstruct $64\times64$ images. The final layer produces 15 output channels, each corresponding to one of the 15 two-dimensional projections of the 6D phase-space distribution at the cathode. These 15 predicted histograms are compared to the corresponding simulated truth histograms using a composite loss function combining Poisson loss, mean absolute error, and cosine similarity.

For stability, the transformer block is frozen during Stage~1 so that the encoder and decoder can first learn a consistent mapping without the additional nonlinearity changing every step. With only a single $x-y$ image, the encoder-decoder alone cannot reasonably reconstruct a 6D phase space; in Stage~1, the transformer adds the missing complexity by reprocessing the same encoded image through its attention layers and producing a richer feature representation that makes a 6D approximation learnable. In Stage~2, we then unfreeze the transformer and train it together with the encoder and decoder, because at this stage it must learn how to combine genuinely different $x-y$ images (coming from different RF-solenoid settings) into a single consistent 6D solution.

\subsection{Loss function}
The loss function is crucial in machine learning as it numerically evaluates the accuracy of a model's predictions and serves as a criterion for determining the direction of learning.
A loss function represents the magnitude of the discrepancy between a predicted value and the correct value. In this study,
we use a loss function which is a sum of Poisson negative log-likelihood term $\mathcal{L}_{\text{Poisson}}$~\cite{Moussaoui2025}, a Mean Absolute Error (MAE) term $\mathcal{L}_{1}$~\cite{Zhao2016}, and a cosine similarity term $\mathcal{L}_{\text{cos}}$~\cite{Luo2018} given by,
\begin{equation}
  \mathcal{L}
  = \mathcal{L}_{\text{Poisson}}
  + \mathcal{L}_{1}
  + \tfrac{1}{2}\,\mathcal{L}_{\text{cos}} ,
\end{equation}
where the three terms are computed over all 15-2D histograms of the 6D phase space. Let $\hat{y}_n$ and $y_n$ denote the predicted and true pixel values, with $n$ indexing every pixel across all histograms and batch elements, and $N$ is the total number of indexed pixels. The loss terms are defined as follows:
\begin{equation}
  \mathcal{L}_{\text{Poisson}}
  = \frac{1}{N}\sum_{n}\bigl(\hat{y}_n - y_n \log \hat{y}_n\bigr),
\end{equation}
\begin{equation}
  \mathcal{L}_{1}
  = \frac{1}{N}\sum_{n} \lvert \hat{y}_n - y_n \rvert,
\end{equation}
and
\begin{equation}
	\mathcal{L}_{\text{cos}}
	= 1 - \frac{\hat{\mathbf{y}}\cdot\mathbf{y}}
	{\lVert \hat{\mathbf{y}}\rVert\,\lVert \mathbf{y}\rVert},
\end{equation}
where $\hat{\mathbf{y}}$ and $\mathbf{y}$ are the predicted and true histograms flattened into vectors. We use three terms because each captures a different aspect of the histograms. The Poisson loss makes the network respect the total number of particles in each bin. The MAE term penalizes the absolute difference in intensity at each pixel and helps the model match the local details of the histograms. The cosine similarity loss is applied to the whole flattened histogram and encourages the predicted and true distributions to have a similar global shape and pattern.

\subsection{Training process}
\begin{figure}[httb]
\centering
  \includegraphics[width=0.75\linewidth]{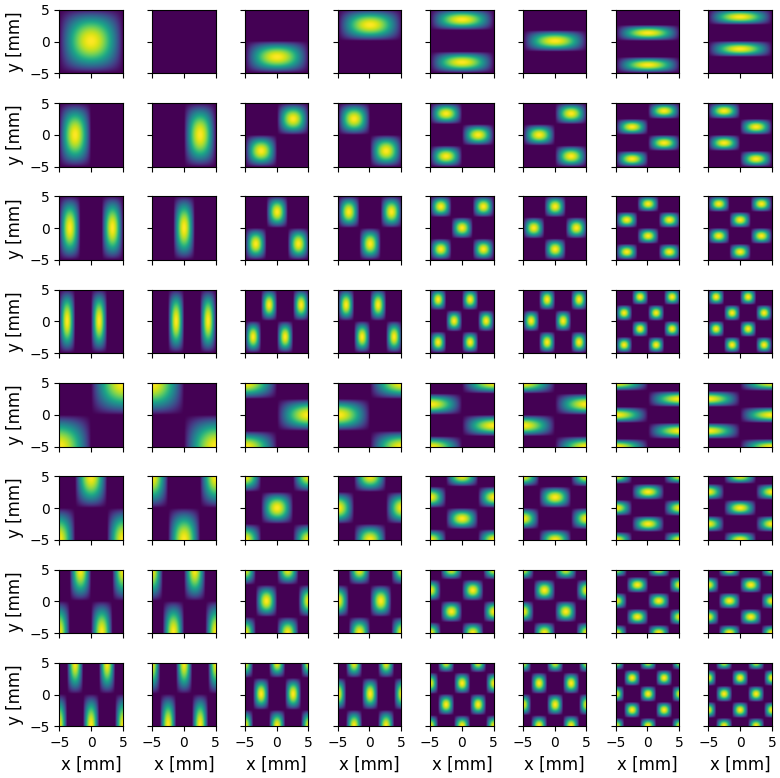}
  \caption{Cathode beam distributions that are generated using Fourier series functions. These provide various beam shapes for the model training. By providing Fourier generated functions we improved the CNN's ability to extrapolate to experimental beam shapes.}
  \label{fig:fourier_dist}
\end{figure}

To train the model of CNN on a wide variety of beam shapes, we generated the cathode $x-y$ distributions using Fourier-type patterns of the form $\sin(nx)\sin(ny)$ and $\cos(nx)\cos(ny)$ with integer frequencies $n=1,\ldots,4$, giving up to four peaks in both $x$ and $y$. When forming histograms, we clip negative values to zero, so to cover all possible peak configurations, we also use $\pi$-shifted variants of these modes. The resulting transverse profiles are shown in Fig.~\ref{fig:fourier_dist}. Only the $\pi$-shifted $\sin(1x)\sin(1y)$ mode (second panel from the top left), which becomes entirely negative and therefore yields a blank histogram after clipping, is excluded from the dataset. The temporal profile is kept Gaussian, and the momentum distribution has isotropic emission angles into a half-sphere with a thermal distribution, consistent with emission from a Cs$_2$Te photocathode~\cite{powell1973}. 

\begin{figure}[httb]
	\centering
	\includegraphics[width=0.8\linewidth]{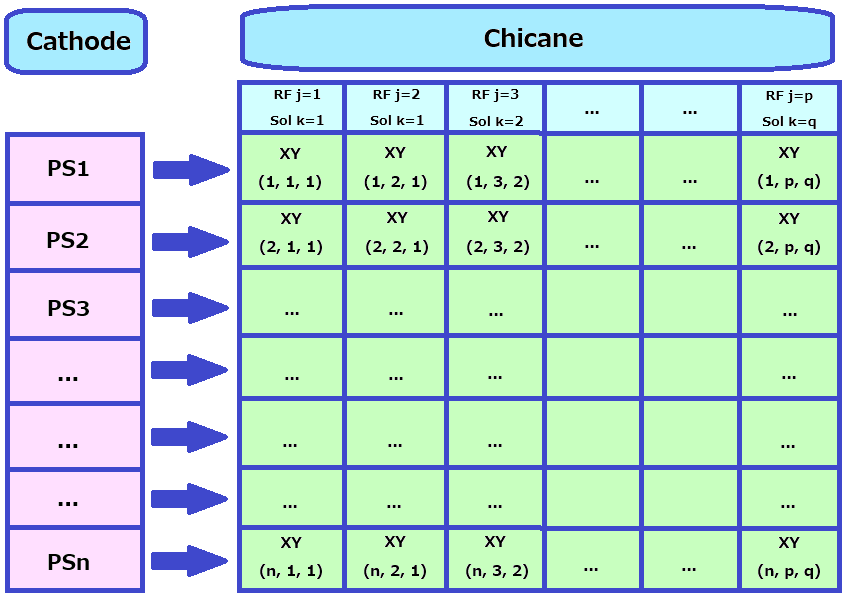}
	\caption{Structure of the dataset used during training. Each row corresponds to a cathode phase space distribution $\mathrm{PS}_i$, and each column corresponds to a $x-y$ image obtained at the chicane by a RF-solenoid combination $(\mathrm{RF}_j,\mathrm{Sol}_k)$. The entry $\mathrm{XY}(i,j,k)$ is the chicane $x-y$ image obtained for that setting.}
	\label{fig:data_structure}
\end{figure}

Figure~\ref{fig:data_structure} summarizes the data structure. Each row corresponds to a single cathode phase space distribution represented by $\mathrm{PS}_i$ ($i=1,\ldots,n$). For each $\mathrm{PS}_i$ we propagate the beam through multiple beamline configurations obtained by combining an RF setting indexed by $j$ ($j=1,\ldots,4$) with a solenoid setting indexed by $k$ ($k=1,\ldots,4$). The resulting chicane image is therefore labeled by three indices: the element $\mathrm{XY}(i,j,k)$ is the chicane $x-y$ image produced when $\mathrm{PS}_i$ is transported with the settings $(\mathrm{RF}_j,\mathrm{Sol}_k)$, where those are RF and solenoid settings, respectively.

In the first stage of the model we fix the RF phase and solenoid field, i.e., selecting one pair $(j,k)$ at a time, working column--by--column. For that fixed configuration, we train on the inverse mapping from each measured chicane image $\mathrm{XY}(i,j,k)$ together with its RF-solenoid setting back to the corresponding 6D cathode phase space $\mathrm{PS}_i$. We then move to a different RF-solenoid combination (a different pair $(j,k)$) and repeat the process, so the network learns how variations in the cathode distribution translate into different chicane images under different fixed beamline configurations. However, a single $x-y$ projection is highly ambiguous: many different 6D distributions can produce the same 2D image.

In Stage~2 we switch perspective and work row--by--row. For each cathode distribution $\mathrm{PS}_i$ we take the collection of images acquired over all scanned RF phases and solenoid fields, $\{(\mathrm{XY}(i,j,k), \mathrm{RF}_j, \mathrm{Sol}_k)\}~(i,j=1,\ldots,4)$, as a multi-view input, and train the network to fuse the information from these views and output a single 6D phase space distribution $\mathrm{PS}_i$ that is consistent with all projected images. Both stages are essential: Stage~1 teaches the model how changes in the cathode distribution modify the measured image for a fixed RF-solenoid configuration, while Stage~2 teaches it to merge images from multiple RF-solenoid settings into a single 6D solution. Furthermore, for each cathode distribution, we varied the temporal profile and the electron kinetic energy as shown in Tab. \ref{tab:initial_beam_cond}.

\begin{table}[httb]
  \centering
  \caption{Beam parameters at cathode used in ASTRA simulations.}
  \label{tab:initial_beam_cond}
  \begin{tabular}{lccc}
    \hline
    Parameter                         & Range          & Step size   & Unit \\ \hline
    Laser pulse duration (FWHM)       & $5.00$--$14.40$ & $4.70$ & ps   \\
    Electron kinetic energy $E_{K}$   & $0.45$--$0.65$ & $0.10$ & eV  \\ \hline
  \end{tabular}
\end{table}

To optimize the model, we check the lowest validation loss over a certain number of training cycles or epochs. The model makes small groups called mini-batches~\cite{Bottou2018} of samples from the whole dataset, which it sees in one update step and uses the Adam optimizer~\cite{Kingma2017} that provides a learning rate~\cite{smith2017} during training. The selection of mini-batch size and learning rate is crucial because it highly influences the model's overall performance, whether it is learning properly from the training samples and can perform well on datasets outside training.

We generate 1700 unique cathode phase space distributions, each associated with 16 chicane $x-y$ images. 
Of the 1700 distributions, 80\% were used as training data and 20\% as test data. 
In Stage~2, the transformer processes 16 different $x-y$ images at the chicane together to reconstruct a single cathode phase space sample, so 
a single cathode phase space sample and corresponding 16 $x-y$ images are treated as one data set.  
The number of data sets in Stage~2 are 1360 for training and 340 for validation. 
In Stage~1,  a cathode phase space sample and one $x-y$ image at the chicane are sent to the transformer; one data set is composed from 
one $x-y$ image at the chicane and a cathode phase space sample. Because there are 16 $x-y$ images at the chicane for a cathode phase space, 
number of the data sets in Stage~1 are 16 times larger that in Stage~2. 
This yields $1360\times16=21760$ training data sets and $340\times16=5440$ validation data sets.

\begin{figure}[httb]
  \centering
  \includegraphics[width=0.48\linewidth]{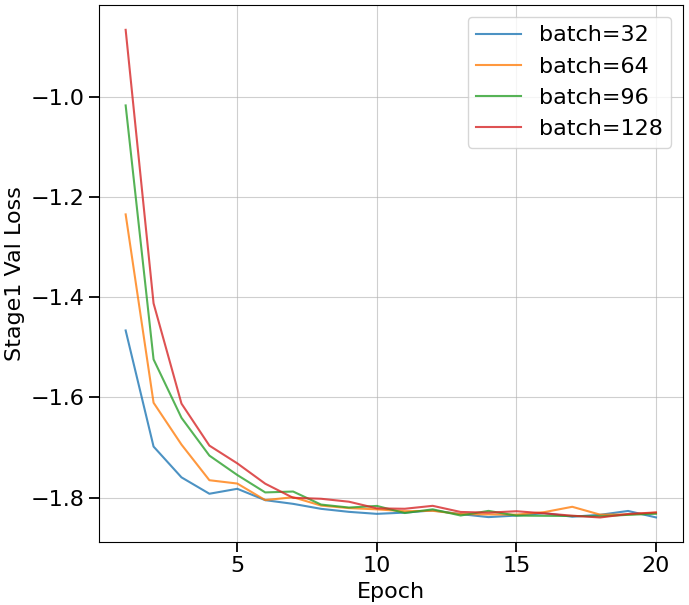}
  \caption{Stage~1 batch size scan: validation loss versus epoch for four different batch sizes: blue, orange, green, and red for batch sizes 32, 64, 96, and 128, respectively.  
  }
  \label{fig:stage1batch}
\end{figure}

The optimization for Stage~1 is shown in Fig.~\ref{fig:stage1batch}. The horizontal axis is the progress of the epoch, and the vertical axis shows the loss function value. 
Several curves for different batch sizes are drawn with different colors: blue, orange, green, and red for batch sizes 32, 64, 96, and 128, respectively. 
Given a fixed learning rate, we observe that for very small batches (32 or 64) each update is based on a small sample and the gradient estimate becomes noisy and slow to converge.
We observe that for larger batches (96 and 128), the validation curve converges well. We therefore select a batch size of 128 for training optimization. 

\begin{figure}[httb]
  \centering
  \includegraphics[width=0.48\linewidth]{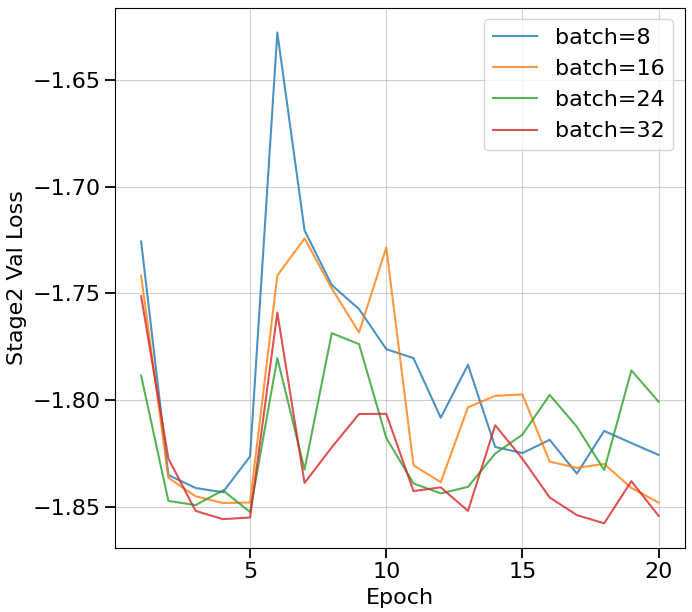}
  \caption{Stage~2 batch size scan: validation loss versus epoch for four different batch sizes: blue, orange, green, and red for batch sizes 8, 16, 24, and 32, respectively. 
  }
  \label{fig:stage2batch}
\end{figure}

Figure~\ref{fig:stage2batch} shows the evolution of the loss function for each epoch for Stage~2, with blue, orange, green, and red now representing batch sizes of 8, 16, 24, and 32, respectively. 
Here, the number of data sets is small compared to Stage~1 and hence we checked the performance on lower batch sizes. For lower batches like 8 or 16, there are noisy gradients and convergence is slow, and the fluctuation decreases on bigger ones, such as 24 and 32. The batch size of 32 was chosen. The spike after epoch 5 is explained in a later paragraph where we talk about the learning rate for Stage~2. It is to be noted that we did the optimization of Stage~2 after optimizing both batch size and learning rate for Stage~1. The tuning of the learning rate started once we obtain the batch size. In Stage~1, the learning is done by the encoder and decoder blocks, and the transformer is kept frozen as described earlier. 

\begin{figure}[httb]
  \centering
  \includegraphics[width=0.48\linewidth]{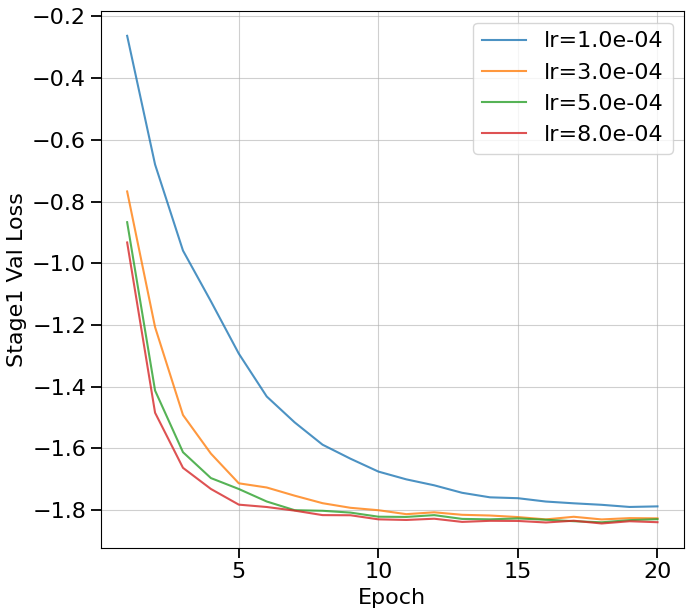}
  \caption{The evolutions of the loss function for Stage~1 for various learning rates are shown, where blue, orange, green, and red lines show the evolution for learning rates $1.0\times 10^{-4}$, $3.0\times 10^{-4}$, $5.0\times 10^{-4}$, and $8.0\times 10^{-4}$, respectively.}
  \label{fig:stage1lr}
\end{figure}

We fix the batch at 128 and then scan over different learning rates, choosing the one that both converges well within the allotted epochs and gives the lowest validation loss. 
Figure~\ref{fig:stage1lr} shows the evolution of the loss function for Stage~1 for various learning rates, where blue, orange, green, and red lines show the evolution for learning rates $1.0\times 10^{-4}$, $3.0\times 10^{-4}$, $5.0\times 10^{-4}$, and $8.0\times 10^{-4}$, respectively. 
According to Fig.~\ref{fig:stage1lr}, low learning rates make the process slow and hence convergence take longer epochs.
We chose a learning rate of $\mathrm{8.0e-4}$ for Stage~1. 

\begin{figure}[httb]
  \centering
  \includegraphics[width=0.48\linewidth]{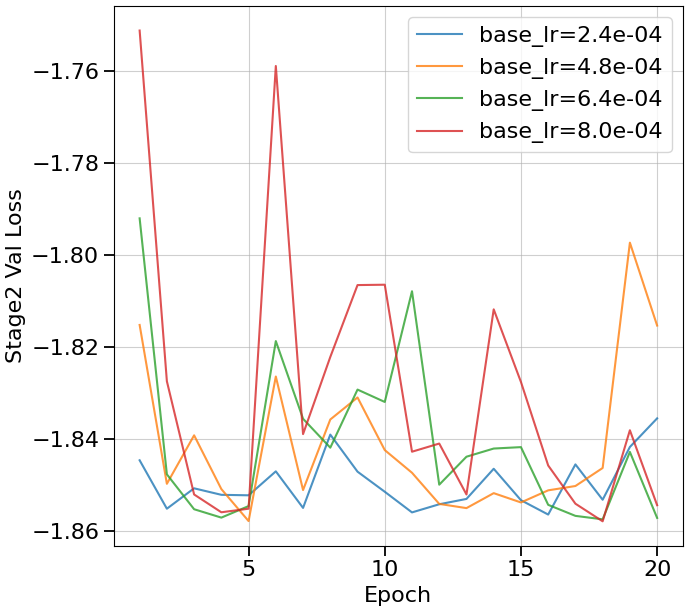}
  \caption{The evolutions of the loss function for Stage~2 for various base learning rates are shown, where blue, orange, green, and red lines show the evolution for learning rates $2.4\times 10^{-4}$, $4.8\times 10^{-4}$, $6.4\times 10^{-4}$, and $8.0\times 10^{-4}$, respectively.}
  \label{fig:stage2lr}
\end{figure}

In Stage~2, the transformer of the model is included as a part of the optimization. We start from the converged Stage-1 weights, then freeze the projection encoder and decoder and train only the transformer for the first five epochs. This lets the transformer learn how to combine the 16 different $x-y$ images without disturbing the single-view mapping. After this warm-up, we unfreeze all three blocks and continue training them together so that the whole network can adapt to the multi-view setting. For clarity, all learning rates are expressed in terms of a single base learning rate. During the warm-up phase, the transformer uses $1.5\times\text{base\_lr}$, and once all blocks are unfrozen, the encoder and decoder use the $0.5\times \text{base\_lr}$ value each, while the transformer uses $1.25\times\text{base\_lr}$. These ratios were chosen from small pilot runs and then kept fixed. 
Figure ~\ref{fig:stage2lr} shows the evolutions of the loss functions in Stage~2 with blue, orange, green and red lines for the various base learning rates, $2.4\times 10^{-4}$, $4.8\times 10^{-4}$, $6.4\times 10^{-4}$, and $8.0\times 10^{-4}$, respectively. 
The base learning rate $6.4\times 10^{-4}$ and $8.0\times 10^{-4}$ showed a good convergence. $8.0\times 10^{-4}$ was chosen as the base learning rate. The spike in the validation loss just after epoch 5 or the warm-up coincides with the unfreezing of the encoder and decoder: at that point, the latent representations and their decoding both start to change, so the transformer is briefly out of sync with the other blocks before the full network re-adapts and the loss decreases again.

\subsection{Cross validation performance}\label{sec:cv_discuss}
\begin{figure}[httb]
	\centering
	\includegraphics[width=0.85\linewidth]{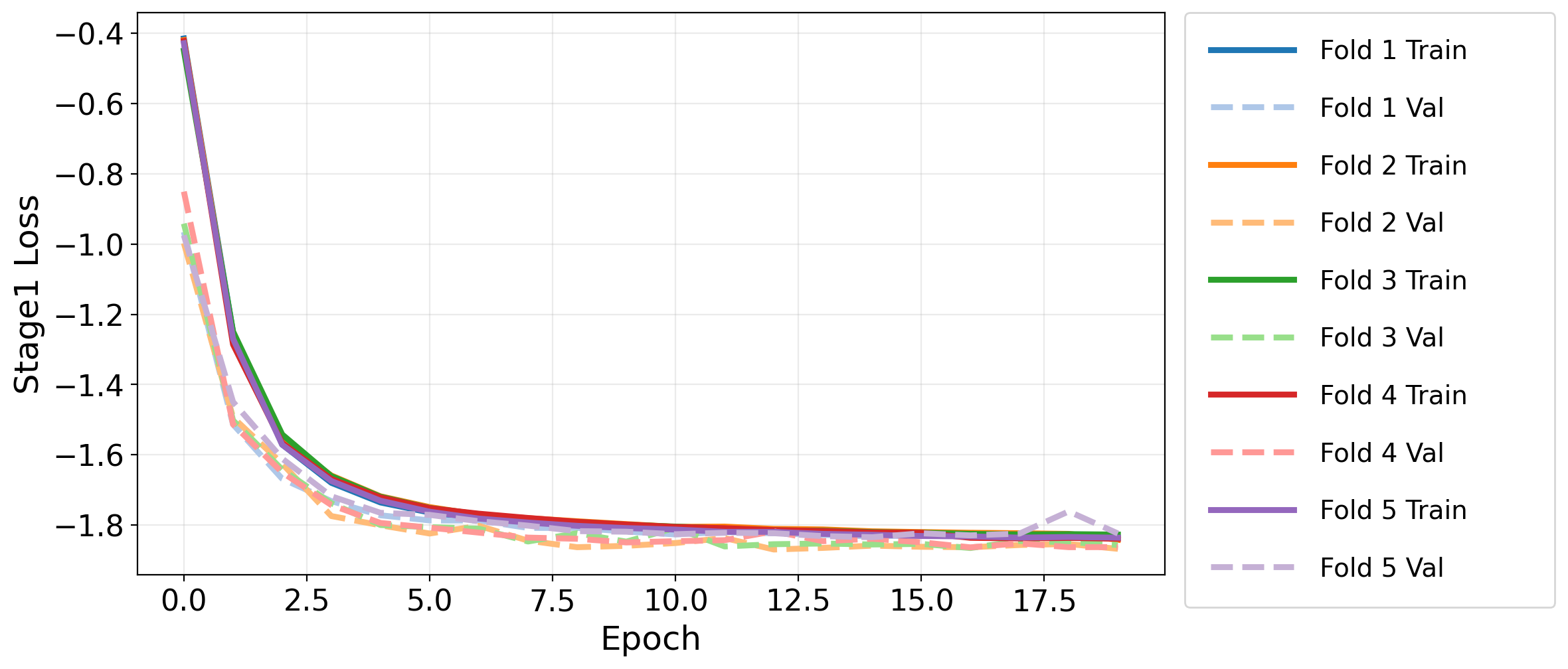}
	\caption{Stage~1 training and validation loss as a function of epoch for each fold in the five-fold cross validation. Fold 1, Fold 2, Fold 3, Fold 4 and Fold 5 are denoted by blue, orange, green, red and purple lines. The solid lines represent the training curves whereas the dotted ones show the validation curves.}
	\label{fig:stage1cv}
\end{figure}

\begin{figure}[httb]
	\centering
	\includegraphics[width=0.85\linewidth]{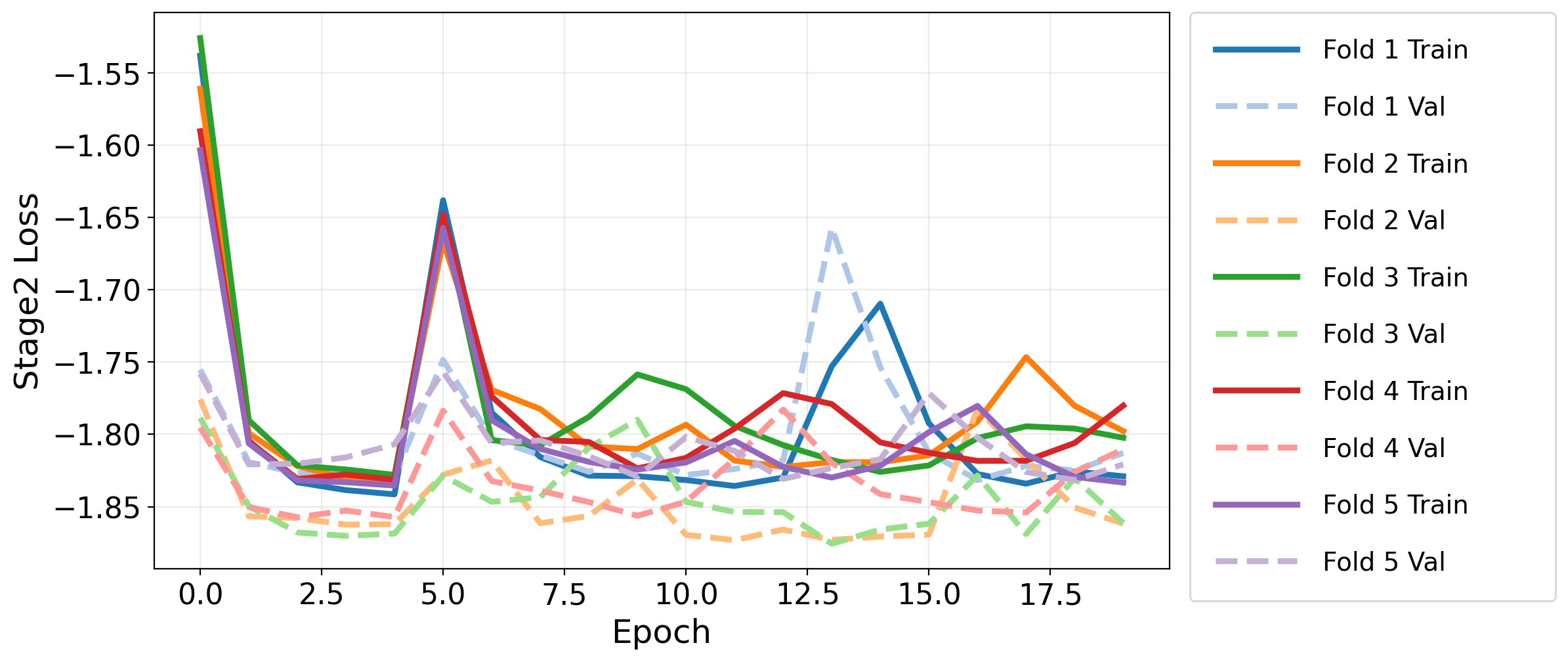}
    \caption{Stage~2 training and validation loss as a function of epoch for each fold in the five-fold cross validation. Fold 1, Fold 2, Fold 3, Fold 4 and Fold 5 are denoted by blue, orange, green, red and purple lines. The solid lines represent the training curves whereas the dotted ones show the validation curves.}
	\label{fig:stage2cv}
\end{figure}
Cross-validation~\cite{Kohavi1995,stone1974}  is a technique used to accurately evaluate and predict the performance of machine learning models and stabilize the model's generalization ability. It involves repeatedly splitting the limited available data into “training” and “test” sets and reusing them. 
This prevents overfitting and helps create reliable models that are not influenced by data bias. 
Here we use five-fold cross validation, where the data are divided into five parts and, in each run, four folds are used for training and the remaining fold for validation; the final performance is then averaged over the five runs. Before cross-validation, we first ran a single training run to determine the optimal batch size and learning rate for each stage. We then performed cross-validation separately for Stage~1 and Stage~2 using their respective datasets (which differ in size), while keeping the stage specific optimal hyperparameters fixed.

Figure~\ref{fig:stage1cv} shows the evolution of the loss function in Stage-1 training with the 5 folds. Solid curves denote training loss and dashed curves denote validation loss, with each fold’s train-validation pair plotted in the same colors: blue, orange, green, red and purple for Fold 1, Fold 2, Fold 3, Fold 4, and Fold 5, respectively.  Some differences are seen among the folds, which arises from the fact that in each fold the subset of data is different and hence the model's performance varies. Figure ~\ref{fig:stage2cv} summarizes the 5-fold performance for Stage~2 in the same manner. Here, folds 2 and 3 show higher differences in training and validation losses than the others, and fold 5 shows the minimum. However, these differences stay within a small range of loss values. To get a consistent result from our two-stage model, we averaged the performance from all folds.

\section{Performance evaluation with synthetic distributions}
\label{sec:synthetic_samples}

To evaluate the performance of the CNN model we constructed, we conducted tests using two methods. 
One method utilized simulation data, while the other employed experimental data from KEK-ATF.

In the simulation-based test, we fed the CNN with 16 real-space distributions obtained from different RF-Solenoid settings, 
derived from a simulated beam distribution on a cathode. 
We then compared the reconstructed distribution on the cathode with the one provided to the simulation.

In the experimental data test, while the complete phase-space distribution on the cathode is unknown, 
it is possible to infer the $x-y$ distribution and time-dependent distribution of the generated beam from the shape and temporal 
distribution of the laser irradiating the photocathode. 
These inferred distributions were compared with the reconstructed phase-space distribution. 
Furthermore, since the momentum-space distribution is determined by the thermal distribution of the cathode, 
an evaluation of its validity was performed.

\subsection{Validation on simulated test sets}

\begin{figure*}[httb]
	\centering
	\includegraphics[width=\textwidth]{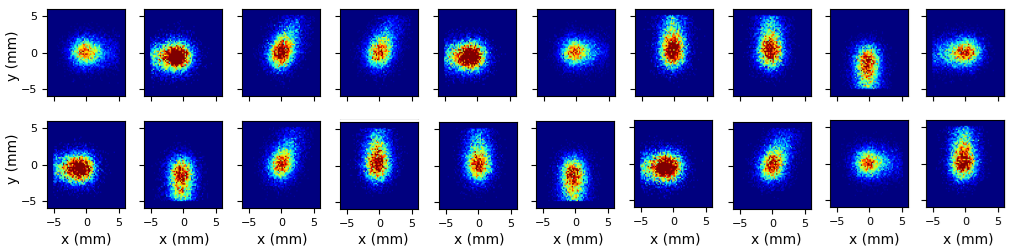}
	\caption{The $x-y$ distributions for the test set are shown. These are the cathode distributions obtained after introducing a tail-like function to Gaussian distribution. The orientation for the tails are taken over $0^\circ$ to $270^\circ$. Thus, we obtained samples that resemble ``comet'' shapes. These test distributions were not included in the training set and are used only to evaluate the CNN’s ability to generalize to unseen beam shapes.}
	\label{fig:testdistributions}
\end{figure*}

To examine the CNN with simulation data,we generated cathode $x-y$ distributions.
To simulate more realistic beam shapes, an asymmetric perturbation was applied and the CNN's response to it was examined.
The distribution $P_\theta(x,y)$ for the examination is given as 
\begin{equation}
	P_\theta(x,y) = (1-w)\,G(x,y) + w\,T_\theta(x,y),
\end{equation}
where $w\in[0,1]$ sets the perturbation strength with $w=0$ giving a pure Gaussian and larger $w$ producing a more pronounced perturbation,
$\theta$ is direction of a tail. $G(x,y)$ is a Gaussian distribution given as
\begin{equation}
	G(x,y)=C_G\exp\!\left[-\frac{(x-x_c)^2}{2\sigma_x^2}-\frac{(y-y_c)^2}{2\sigma_y^2}\right],
\end{equation}
where $C_G$ is a normalization constant chosen such that $\sum_{x,y} G(x,y)=1$, $(x_c,y_c)$ is the core center, and $\sigma_x$ and $\sigma_y$ set the RMS widths of the Gaussian core in $x$ and $y$, respectively. 
$T_\theta(x,y)$ is the tail function given by convolution as
\begin{equation}
	T_\theta(x,y) = (G* F_\theta)(x,y),
\end{equation}
where $F_\theta$ is defined as
\begin{equation}
	F_\theta(t,s)=
	\begin{cases}
		C_F\exp(-t/L)\,\exp\!\left(-\dfrac{s^2}{2\sigma_t^2}\right), & t\ge 0,\\
		0, & t<0,
	\end{cases}
\end{equation}
where $t$ is the coordinate along the tail direction and $s$ is the coordinate perpendicular to the tail as
\begin{equation}
	t = x\cos\theta + y\sin\theta, \qquad
	s = -x\sin\theta + y\cos\theta ,
\end{equation}
and $C_F$ is a normalization constant for which $\sum_{t,s}F_\theta(t,s)=1$, $L$ is the tail decay length along $t$ and $\sigma_t$ controls the transverse thickness of the tail in the $s$ direction. 
The condition $t\ge 0$ enforces a forward tail. The tail component is then obtained by convolution of the Gaussian and tail functions,
The 20 test samples of $x-y$ distribution were made as shown in Fig.~\ref{fig:testdistributions}, with tail directions varied from $0^\circ$ to $270^\circ$.

It should be noted that the training was performed only with the Fourier images and that the ``comet'' distributions were excluded from the training data. The model therefore had no prior knowledge of these shapes, and therefore successful reconstruction denotes the ability to extrapolate beyond the distributions provided for training.  

\begin{figure*}[httb]
	\centering
	\includegraphics[width=\textwidth]{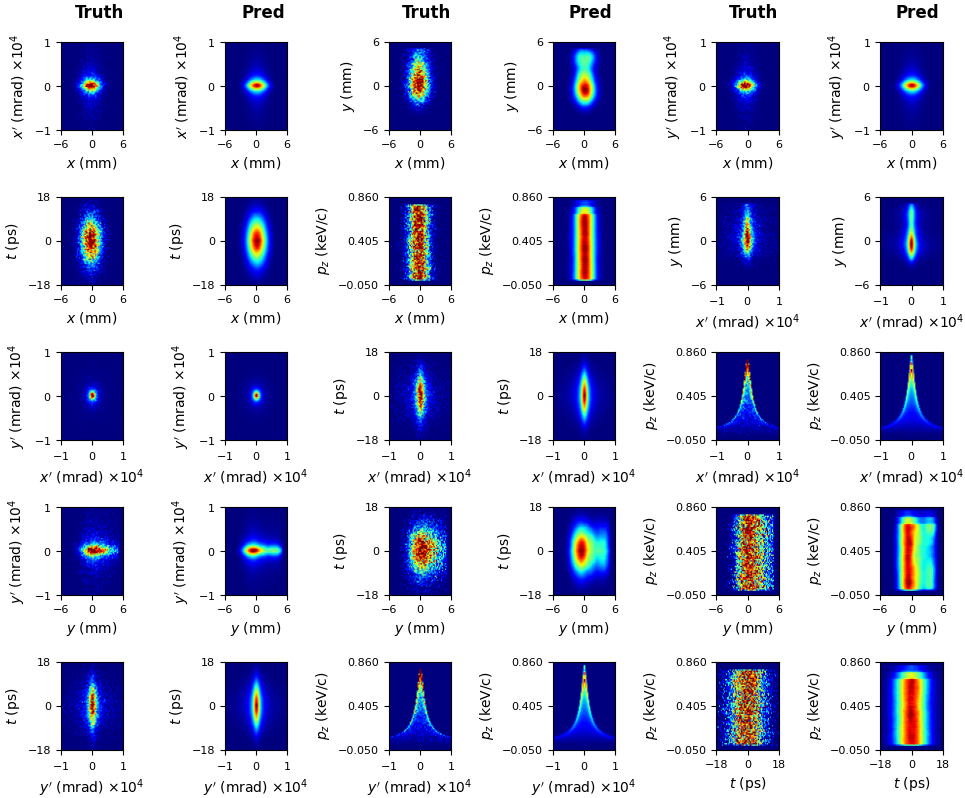}
	\caption{For test sample 1, the comparison of ASTRA simulation and CNN predicted results for the 6D phase space are shown side by side for all combinations of the 6D phase space coordinates. “Truth” is the ASTRA and “Pred” is the CNN reconstruction.}
	\label{fig:resultstest2}
\end{figure*}

\begin{figure*}[httb]
	\centering
	\includegraphics[width=\textwidth]{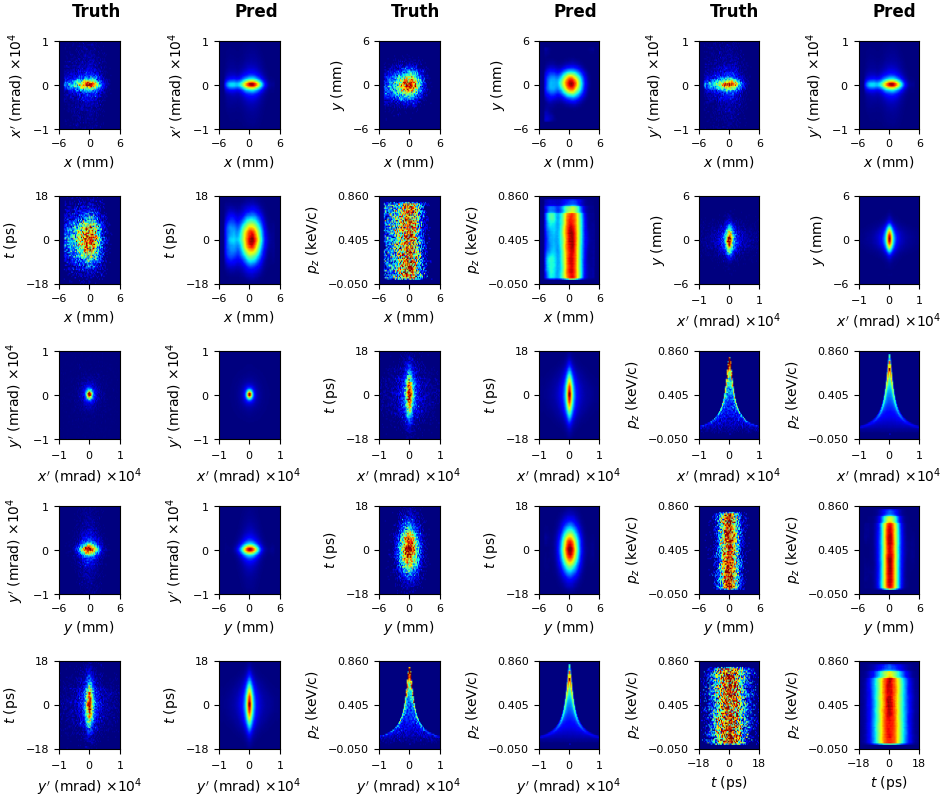}
	\caption{For test sample 2, the comparison of ASTRA simulation and CNN predicted results for the 6D phase space are shown side by side for all combinations of the 6D phase space coordinates. “Truth” is the ASTRA and “Pred” is the CNN reconstruction.}
	\label{fig:resultstest1}
\end{figure*}

For each $x-y$ distribution on the cathodes, the model creates a corresponding six-dimensional phase space. 
Starting from the initial distributions, we generate 16 $x-y$ images at the chicane using different combinations of RF-Solenoid settings and simulating their trajectories with the ASTRA simulation code. 
The 16 generated $x-y$ images are processed by the CNN model and the initial phase-space distributions on the cathode are reconstructed. 
We then compare the reconstructed distributions with the corresponding truth distributions used in the simulation.
The truth and reconstructed images of the 6D phase space for two representative test samples are shown in Figs.~\ref{fig:resultstest2} and~\ref{fig:resultstest1}, respectively. 
For the first sample shown in Fig. ~\ref{fig:resultstest2}, in the transverse planes $x-x'$, $x-y$, $x-y'$, $x'-y$, $x'-y'$, and $y-y'$, the model reproduces well both the Gaussian core and the elongated tail. 
In particular, in $x-y$, $x'-y$, and $y-y'$, the tail is clearly visible in the reconstruction with the correct direction and approximate length. The planes involving the temporal coordinate, $x-t$, $y-t$, $x'-t$, and $y'-t$, also show good agreement: the bunch length and the slight tilt of the distributions are recovered. In the momentum-related projections $x-p_z$, $y-p_z$, $x'-p_z$, $y'-p_z$, and $t-p_z$, the main $p_z$ peak and its narrow tail appear in both truth and prediction. The $t-p_z$ panel in particular shows that the model captures the dominant time-energy correlation, while the very sharp features at the tip of the tail are slightly smoothed. Overall, the core, tail, and associated correlations in $x$, $y$, $t$, and $p_z$ are all reconstructed consistently by the CNN model.
For the second synthetic test beam, shown in Fig.~\ref{fig:resultstest1}, the reconstructed phase-space panels again reproduce the Gaussian cores and skewed tails seen in the truth, similar to the first test case. The main correlations in the $x$, $y$, $t$, and $p_z$ projections are preserved, with only small differences appearing in the very low-density regions.

To quantify the agreement between the reconstructed and reference distributions in each 2D panel, we compute reduced chi-square ($\chi^2_\mathrm{red}$)~\cite{errorbook1992},
\begin{equation}\label{eq:chisq}
\chi^2 = \frac{1}{N_\mathrm{bins}}\sum_{i,j} \frac{\bigl(P_{ij} - T_{ij}\bigr)^2}{T_{ij}},
\end{equation}
where $T_{ij}$ and $P_{ij}$ are the bins of the truth and predicted histograms. Thus $\chi^2_\mathrm{red}$ compares both the shapes and intensities of the distributions.

\begin{figure}[httb]
	\centering
	\includegraphics[width=0.4\linewidth]{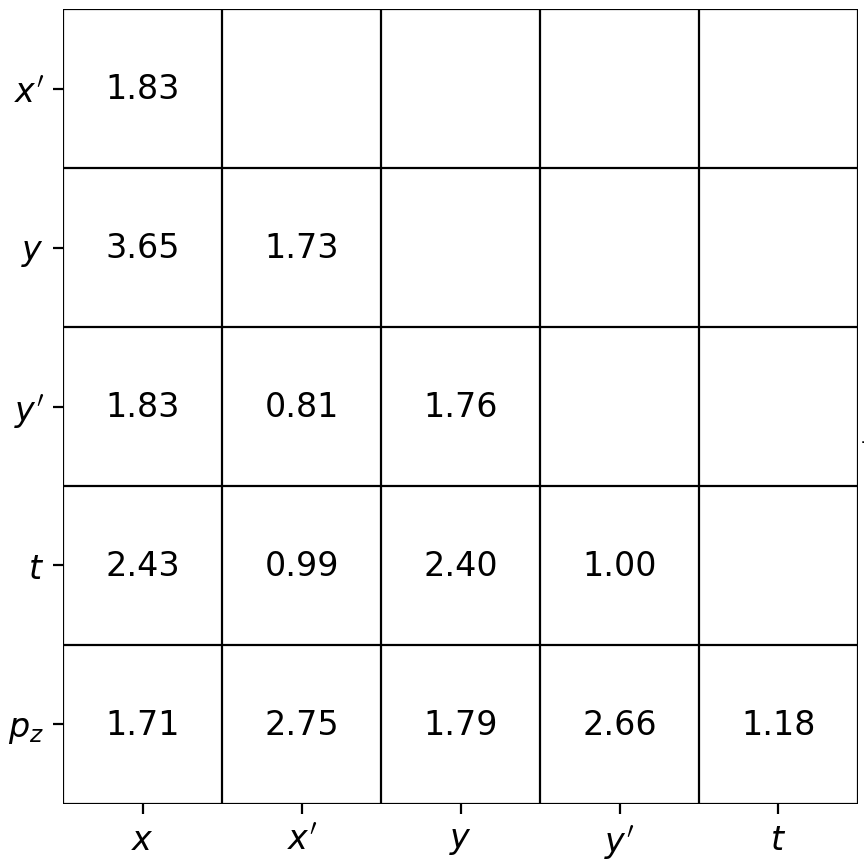}
	\caption{Reduced chi-square ($\chi^2_\mathrm{red}$) values for each combination averaged over 20 test beams.}
	\label{fig:chisqrresults}
\end{figure}

Figure~\ref{fig:chisqrresults} shows the average reduced $\chi^2_\mathrm{red}$ for all 15 phase space planes, obtained by averaging over the 20 test distributions. 
It shows  good quantitative agreements between the reconstructed and truth images.

\section{Experimental Demonstration}
\label{sec:experiment}
With the model sufficiently trained on synthetic beam distributions, we next apply it to a real-world accelerator to demonstrate its performance on real-world beam data. As a suitable test case, we identify the KEK-ATF injector at the KEK Tsukuba campus. Consistent with the needs of the algorithm, the ATF injector includes a phosphor screen within the chicane, allowing observation of changes in beam spread and therefore changes induced in the longitudinal phase-space distribution by changes in the RF and solenoid parameters. With a relatively simple setup, we are able to experimentally demonstrate the feasibility and efficacy of the CNN technique.

\subsection{The KEK-ATF injector}
\label{sec:beam-model}

The KEK-ATF is focused on R\&D work for the ILC and other linear collider projects. It consists of an electron RF gun, a $1.3\,\mathrm{GeV}$ linac, a 1.3\,GeV damping ring, and a beam extraction region for a final focus experiment~\cite{ATF2025}. The linac is composed of S-band normal conducting cavities and operates with an average accelerating gradient of about $25\,\mathrm{MV/m}$ and delivers electron beams at $1.3\,\mathrm{GeV}$. After the beam is stored in the damping ring, the horizontal and vertical emittances are reduced to the order of $10^{-9}\,\mathrm{m\cdot rad}$ and $10^{-12}\,\mathrm{m\cdot rad}$ by radiation damping, respectively. With these ultra-low emittances, the typical transverse beam size has already reached the nanometer level required for ILC operation~\cite{Okugi2016}.

\begin{figure}[httb]
	\centering
	\includegraphics[width=0.7\linewidth]{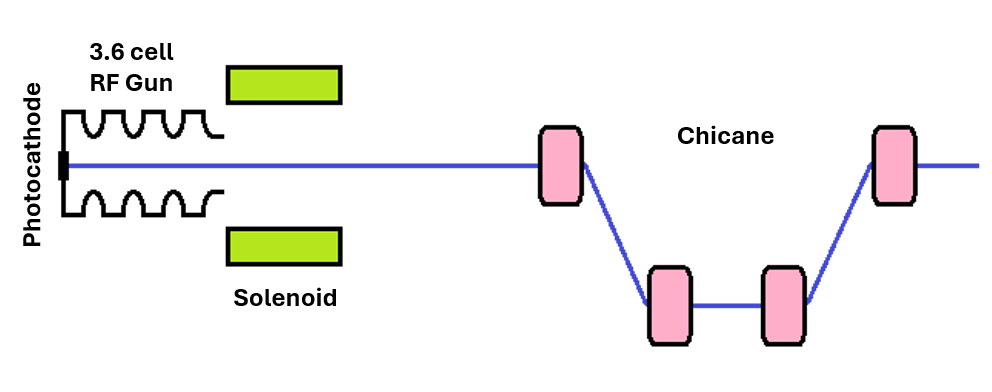}
	\caption{Schematic drawing of KEK-ATF injector which consists of a S-band RF gun, solenoid magnet, and chicane. This beamline is simulated by the ASTRA to generate datasets for the CNN.}
	\label{fig:atf_beamline}
\end{figure}

The portion of the ATF beamline relevant to our study is shown in a schematic in Fig.~\ref{fig:atf_beamline}. It begins with a 2856\,MHz S-band RF gun equipped with 3.6 cell normal conducting cavity that generates electron beams from a Cs$_{2}$Te photocathode illuminated with a UV laser. The RF gun operates with a peak accelerating gradient of about $65\,\mathrm{MV/m}$, producing electron bunches with an energy of approximately $6\,\mathrm{MeV}$ at its exit. The laser pulse has a wavelength of $266\,\mathrm{nm}$ and a repetition rate of $3.125\,\mathrm{Hz}$~\cite{Popov2025}. Immediately after the RF gun, the beam passes through a solenoid, a drift space, and finally a chicane region consisting of four dipoles, each bending the beam by $22.7^\circ$ in the horizontal plane, resulting in a net orbit offset of about $80\,\mathrm{mm}$ in the $x$ direction.

Three sets of the 16 $x-y$ beam images were measured in the chicane region of the KEK-ATF facility, varying RF-Gun phase and solenoid field.
At the KEK-ATF, the laser spot size is modified with a telescopic beam expander system consisting of two convex lenses, where the input laser spot size is expanded or contracted according to the change in focal length of the system. Focal length changes are performed by moving the downstream convex lens, referred to as the zoom lens, while keeping the upstream lens fixed. During our experiment, we changed the optics by adjusting the zoom lens position to three settings: $-1000\,\mu\mathrm{m}$, $-1300\,\mu\mathrm{m}$, and $-1150\,\mu\mathrm{m}$, to produce three different laser spot sizes on the cathode surface. The laser spot size was then measured with a CCD camera, and the resulting images were measured with a screen monitor at the center of the chicane.

Modulation of the RF phase changes the electric field at the cathode surface during beam emission. As a result of the Schottky effect, in which the effective cathode work function is lowered in the presence of increased electric fields, the emission current (beam charge) is altered when the RF phase is changed. The measured beam charge is therefore different for beam images taken at different RF phase values.

\begin{figure}[httb]
  \centering
  \includegraphics[width=0.6\linewidth]{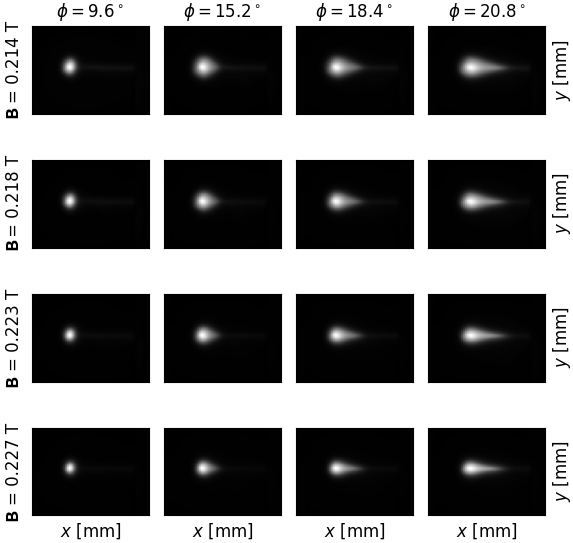}
  \caption{Measured beam images from KEK-ATF, taken on the screen in the chicane section. Each image corresponds to a different combination of RF phase and solenoid peak field. These were obtained by keeping the zoom lens position at $-1150\,\mu\mathrm{m}$.}
  \label{fig:exptproj3}
\end{figure}
For a fixed laser spot size, we varied the RF phase and the solenoid peak field across four settings each. As a result, this procedure generated 16 $x-y$ images at the chicane, which we measured by the beam monitor. This measurement was repeated for all three laser spot sizes and it took about 20 minutes to perform the whole experiment.
The images in Fig.~\ref{fig:exptproj3} were measured at the chicane while keeping the zoom lens position in the laser optics system at $-1150\,\mu\mathrm{m}$. The beam charge, observed by the ICT showed a variation of $0.18$--$0.35\,\mathrm{nC}$ as the RF phase was scanned over $9.6^\circ\text{--}20.6^\circ$. This is due to the rise in photoelectric emission at cathode by Schottky effect with higher RF phases. In parallel, the solenoid field was also varied in the range of $0.214$--$0.227\,\mathrm{T}$. From left to right, the increase in the beam tail with higher RF phase shows the rise of energy spread, and a small decrease in beam size from top to bottom indicates the higher focusing effect of the solenoid field. Similar measurements were also carried out at zoom lens positions of $-1300\,\mu\mathrm{m}$ and $-1000\,\mu\mathrm{m}$, and the corresponding images at chicane are provided in the appendix, where the beam charge was varied over $0.18$--$0.20\,\mathrm{nC}$ and $0.3$--$0.7\,\mathrm{nC}$, respectively, while the RF phase was scanned.

\begin{figure*}[httb]
  \centering
  \includegraphics[width=\textwidth]{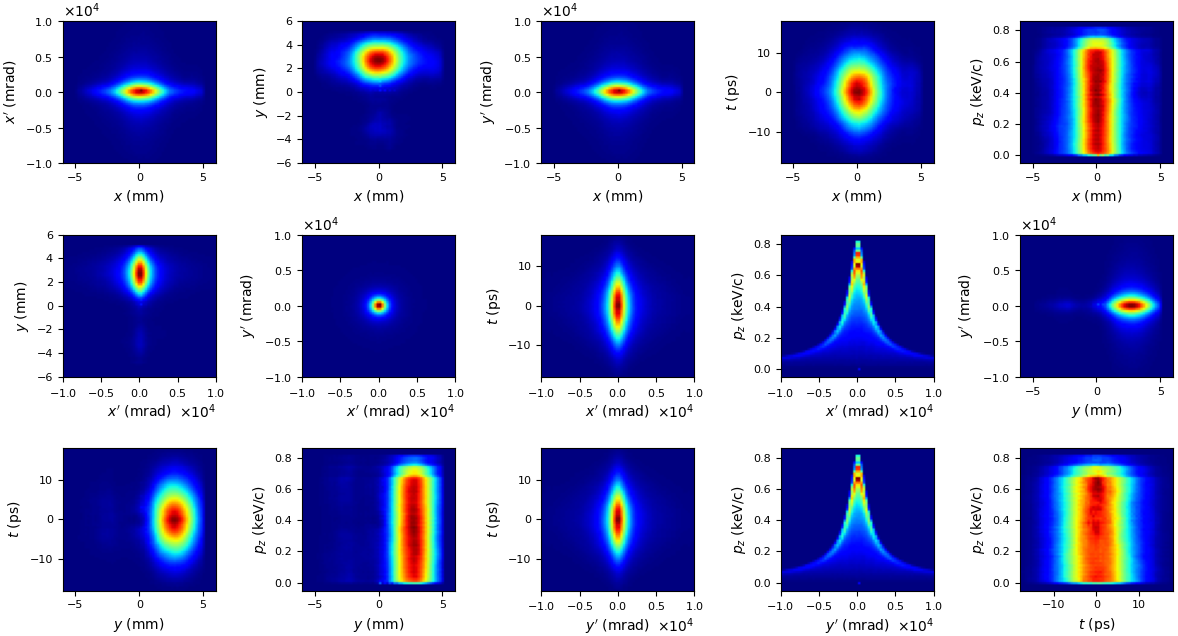}
  \caption{The experimental 6D phase space obtained at zoom lens position of $-1150\,\mu\mathrm{m}$. This result is generated by the CNN from the chicane images shown in Fig.~\ref{fig:exptproj3}.}
  \label{fig:6dimageexpt3}
\end{figure*}

\begin{figure*}[h!]
  \centering
  \includegraphics[width=\textwidth]{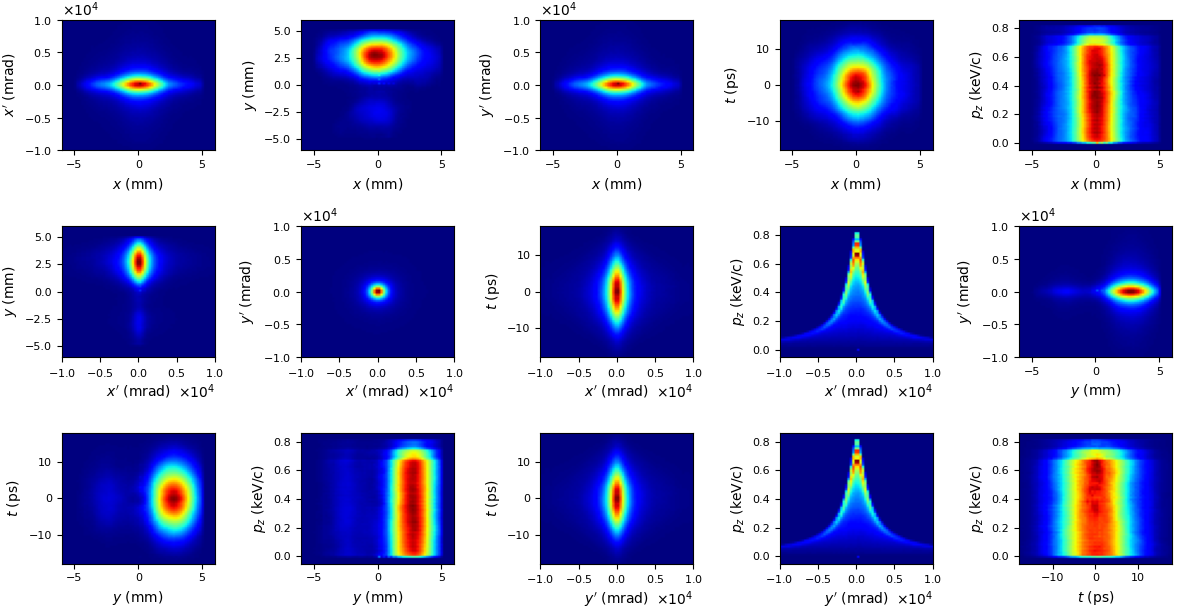}
  \caption{The experimental 6D phase space obtained at zoom lens position of $-1300\,\mu\mathrm{m}$. The CNN used the chicane images shown in Fig.~\ref{fig:exptproj2} as inputs in this case.}
  \label{fig:6dimageexpt2}
\end{figure*}

\begin{figure*}[h!]
  \centering
  \includegraphics[width=\textwidth]{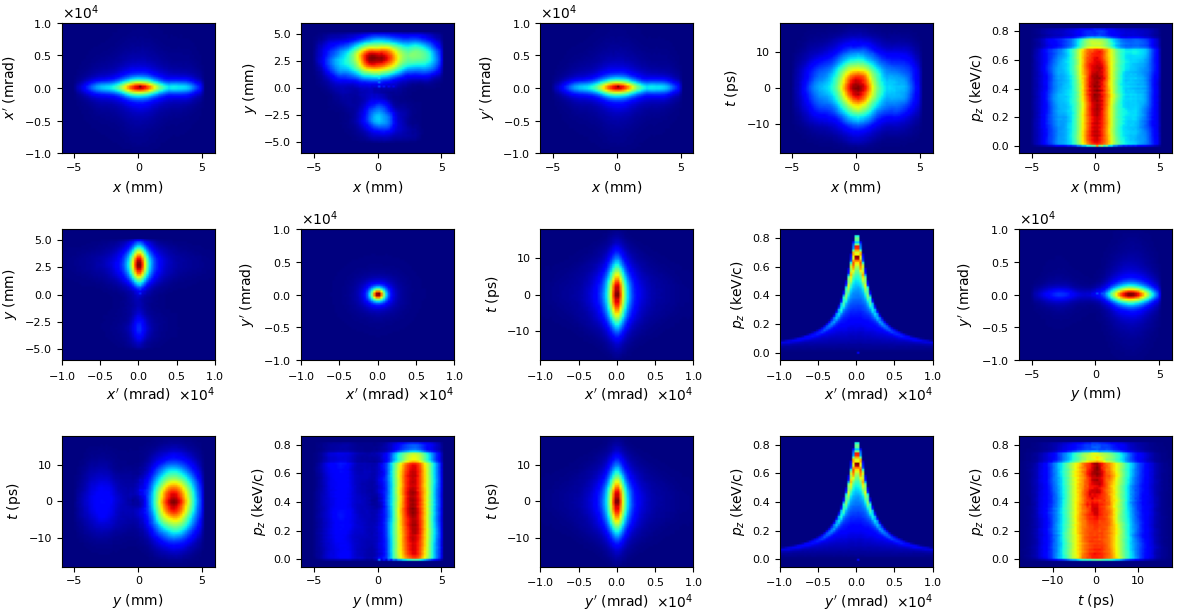}
  \caption{The experimental 6D phase space obtained at zoom lens position of $-1000\,\mu\mathrm{m}$. In this reconstruction, we feed the CNN with the input images shown in Fig.~\ref{fig:exptproj1}.}
  \label{fig:6dimageexpt1}
\end{figure*}

\subsection{Experimental Results}

After obtaining the experimental images, these are used directly as inputs to the pre-trained CNN to generate their respective 6D phase space distributions at the cathode surface. Figures~\ref{fig:6dimageexpt3}-~\ref{fig:6dimageexpt1} show the predicted result of 6D phase spaces corresponding to three zoom lens positions. Since the laser spot sizes at the cathode were different due to the lens positions, the resulting 6D phase spaces also showed slightly different spreads in some planes, such as $x-y$, $x-p_z$, $x-t$, and $x-x'$.

We evaluated the FWHM of the six reconstructed variables $(x, x', y, y', t, p_z)$ across the 15 two-dimensional phase space planes of the reconstructed 6D distribution, and then averaged these values to obtain one representative FWHM per variable. The resulting FWHM values are listed in Table~\ref{tab:fwhm_values} and their comparison with the experimental references are shown. For the transverse beam sizes at the cathode, we compared the CNN generated results to the measurement values of UV laser spot sizes at the above mentioned zoom lens positions. To compare the longitudinal bunch length, the ATF streak-camera measurement is used. The FWHM values of $x$ and $y$ are in the range of a few millimeters, and the $t$ profile shows values in a few picoseconds, approximately consistent with experimental reference values.

\begin{table}[htbp]
	\centering
	\caption{Full width at half maximum (FWHM) comparison between the CNN reconstruction and available experimental references for three zoom lens positions, $-1000\,\mu\mathrm{m}$, $-1150\,\mu\mathrm{m}$, and $-1300\,\mu\mathrm{m}$ are shown. Experimental references for the UV laser spot size at the cathode in $x$ and $y$ are denoted by ``Expt.'', while ``Streak Ref.'' refers to the streak-camera reference for the bunch length $t$. Here, a dash indicates that no experimental reference is available.}
	\label{tab:fwhm_values}
	\begin{tabular}{l|cc|cc|cc|c|c}
		\hline
		& \multicolumn{2}{c|}{Lens: $-1000\,\mu\mathrm{m}$} 
		& \multicolumn{2}{c|}{Lens: $-1150\,\mu\mathrm{m}$} 
		& \multicolumn{2}{c|}{Lens: $-1300\,\mu\mathrm{m}$} 
		& \multicolumn{1}{c|}{Streak} & \\
		Variable 
		& Recon. & Expt.
		& Recon. & Expt.
		& Recon. & Expt.
		& Ref. & Unit \\
		\hline
		$x$   & 3.08   & 2.91  & 2.81   & 2.24 & 3.04   & 1.69 & --   & mm   \\
		$y$   & 2.59   & 4.06  & 2.59   & 2.89 & 2.62   & 1.94 & --   & mm   \\
		$x'$  & 1.56e3 & --    & 1.56e3 & --    & 1.56e3 & --    & --   & mrad \\
		$y'$  & 1.56e3 & --    & 1.56e3 & --    & 1.56e3 & --    & --   & mrad \\
		$p_z$ & 0.71  & --    & 0.71  & --    & 0.71  & --    & --   & keV/c \\
		$t$   & 12.90   & --    & 13.10   & --    & 13.20   & --    & 10.00 & ps   \\
		\hline
	\end{tabular}
\end{table}

{Reconstructed values of the momentum phase-space variables are difficult to compare directly with experiment, but they are assumed to be reasonable based on simulation. They are, however, identical in both $x'$ and $y'$ for all three data sets; likewise, $p_z$ values are identical across all three data sets. We hypothesize this to be the result of a finite number of momentum-space distributions used in training, creating a small number of solutions to which the model converges. Future deployment of the model will include improved momentum phase-space coverage in training to combat this tendency.

\section{Summary of 6D Phase Space Reconstruction }
\label{sec:discussion}

We developed a novel approach for a pseudo 6D phase space reconstruction using the two-stage CNN model. It requires only sixteen real-space beam images on screen, which can be easily obtained from accelerator facilities. In the future, connecting the model to the beam monitor would let it automatically collect images and reconstruct the 6D phase space in real time. In general, the model aims at solving the beam dynamics in two steps: in the first it learns the variation of an upstream 6D beam and its impact on the projected 2D beam at the measurement point, given the beamline configuration with solenoid and RF phase remains constant; in the second, it understands the evolution of a constant 6D beam to different 2D projected images under various RF-solenoid configurations. 

Using the two-step strategy provides the necessary complexity for the model to solve the 6D phase space distribution. We have seen that simply using one stage and increasing the neural network layers or neurons is not sufficient as the mechanisms in each stage are different. Although the Stage~2 is used to combine the real space images to get the phase space, if it is used without the Stage~1, the CNN usually generates an average estimate by combining all the images. Thus, the Stage~1 training makes the model overcome this tendency of skipping important details in the beam distributions. Both stages are therefore needed to make a CNN capable of handling complex beam structures. 

The usage of the CNN helped us to extract the information from beam images, such as beam shape, size, position, etc, as the feature maps. When we combined the image latent with that of RF and solenoid settings, we provided twice as many weights to the latter. This reflects the fact that RF-solenoids are mostly responsible to rotate the phase space in transverse and longitudinal directions. Hence, in other accelerator facilities, if the tuning parameters are different, such as quadrupoles or transverse deflecting cavities, they can follow the same rule to intensify the effect of these components, which are responsible for rotating the phase spaces. In this work, we trained on synthetic cathode distributions constructed from Fourier-series shapes, but in other facilities the same framework could instead be trained on simulated images at other positions along the beamline.

Table~\ref{tab:fwhm_longtable} shows the comparison of FWHM values from prediction ($f_p$) and truth ($f_t$) for each variable of the 6D distribution, obtained from all test distributions. The predicted ones correspond to CNN reconstruction, and the truths are from ASTRA simulations. The ratio, $f_p/f_t$, of these parameters lies near unity, showing good agreement with the reconstruction. However, there are some values that showed high deviations with ratios such as 0.64 in the $y$ axis of sample 20, and 1.84 in the $x$ axis of sample 15.  

However, the other variables in the above mentioned samples showed better agreement with values $0.88\text{--}1.10$ respectively. Given the early stage of development, the model therefore shows overall consistent performance in reconstructing the beam shapes. An important aspect of this study is that we do not require differentiable simulations, so the method can be used directly with standard accelerator codes. All training for this study was carried out on a single NVIDIA RTX~A400 GPU, which is readily available and affordable.

As a data-driven approach, the model has some limitations and several ways of improving its ability and accuracy. The CNN can only perform reliably within the range of beam conditions, machine settings, and Fourier-series orders represented in the training data, so its predictions may degrade outside these ranges. Further, despite training and validation loss parameters indicating a lack of overfitting, the model is still vulnerable to imperfect coverage of the phase space parameters to be reconstructed and can fall into a set of ``preferred'' values.
Additionally, the hyperparameter scans presented here were limited to a narrow range, and further improvements may be possible with a broader scan. However, the model showed its ability to exterpolate on test beam shapes which were outside the training. Therefore, the reconstruction performance can be expected to improve in the future if we incorporate higher order frequencies in the Fourier patterns and more variations on the hyper-parameters.

\section{Conclusion}
\label{sec:conclusion}

We demonstrated the pseudo 6D phase space reconstruction by the two-stage CNN model complementary to GPSR-type approaches. In contrast to GPSR, it works with existing non-differentiable simulation codes and delivers reconstructions in under a minute on an NVIDIA RTX~A400 GPU. Its performance is broadly consistent, with reduced chi-square values between the reconstructed and true images
mostly in the range $1.0 \lesssim \chi^2_\mathrm{red} \lesssim 2.0$. The FWHM ratio test on the 20 synthetic beams further shows that $f_p/f_t$ stays close to unity for most variables. The CNN reconstructions from experimental inputs yield transverse beam sizes of $2.6$--$3.0\,\mathrm{mm}$ and bunch lengths of $13.0$--$13.2\,\mathrm{ps}$. For comparison, we use the FWHM of UV laser spot size at the cathode and the streak-camera bunch length reference, which indicates $2.0$--$4.0\,\mathrm{mm}$ transversely and approximately $10.0\,\mathrm{ps}$ temporally. Overall, the reconstruction agrees with these references at the level of a few millimeters in the transverse plane and a few picoseconds in time. The slight differences observed in some results should be understood through future research.
Refinement of the model are likely to improve agreement between reconstructed beam values and experimental measurements. 
Improvements are possible through more realistic modeling of the cathode, improvements in simulation, optimization of hyperparameters, and the addition of higher-order Fourier terms in synthetic beams. However, even at this stage, due to its low computational load, the method should be useful for beam diagnostics in accelerators where beam quality is important, such as synchrotron radiation facilities, free electron lasers, and colliders.

\section*{Acknowledgment}
This work is partly supported by JSPS KAKENHI Grant Numbers JP20H01934.


%

\vspace{0.2cm}
\noindent


\let\doi\relax


\bibliographystyle{ptephy}
\bibliography{references}   

\clearpage

\appendix

\section{Appendix: Experimental images with lens positions $-1300\,\mu\mathrm{m}$ and $-1000\,\mu\mathrm{m}$}

We used three different zoom lens positions for generating different laser spot sizes at the cathode. In the Sec.~\ref{sec:experiment}, the measurement images at chicane corresponding to lens position of $-1150\,\mu\mathrm{m}$ are given. Here, the measurements for the other two lens positions are shown. Figures~\ref{fig:exptproj2} and~\ref{fig:exptproj1} are obtained with zoom lens positions of $-1300\,\mu\mathrm{m}$ and $-1000\,\mu\mathrm{m}$ respectively.

\begin{figure}[httb]
	\centering
	\includegraphics[width=0.6\linewidth]{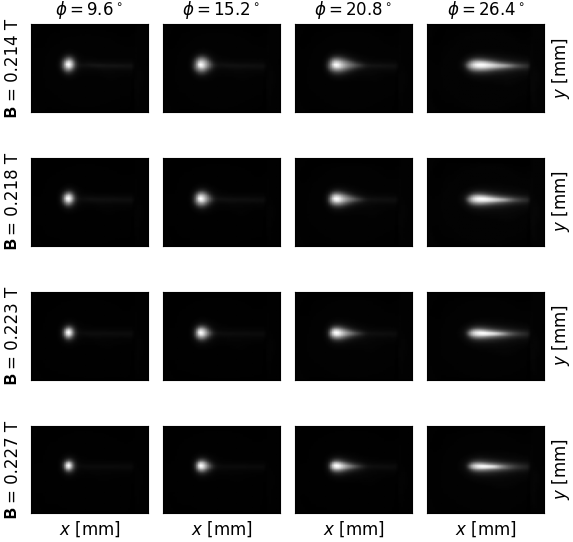}
	\caption{Measured beam images from KEK-ATF, taken on the screen in the chicane section with different combination of RF phase and solenoid peak field. The zoom lens position was kept at $-1300\,\mu\mathrm{m}$.}
	\label{fig:exptproj2}
\end{figure}

\begin{figure}[httb]
	\centering
	\includegraphics[width=0.6\linewidth]{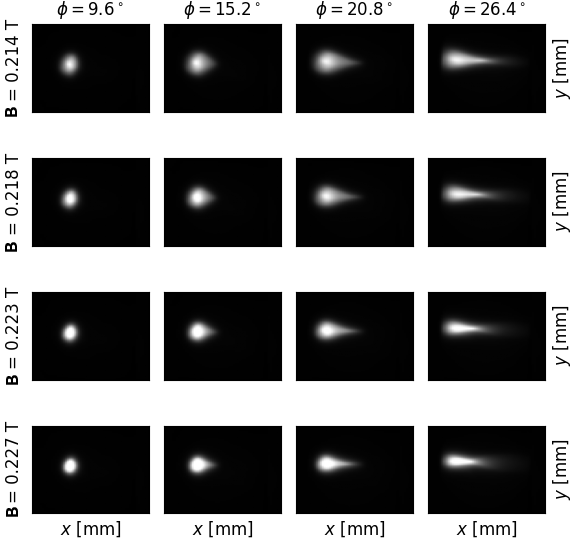}
	\caption{Measured beam images from KEK-ATF, taken on the screen in the chicane section with different combination of RF phase and solenoid peak field. The zoom lens position was kept at $-1000\,\mu\mathrm{m}$.}
	\label{fig:exptproj1}
\end{figure}  

\clearpage

\section{Appendix: CNN layers and their working principles}

Here we describe the main features of Convolutional Neural Networks (CNN), following the treatment given in Ref.~\cite{alzubaidi2021}.
Unlike other neural networks, CNNs take 2D images as inputs and apply convolutional filters to extract spatial features. This approach efficiently extracts important features from the images that can be used to solve pattern recognition problems~\cite{gu2018,yao2019,simonyan2014}. The CNN mainly consists of Convolution layers, Pooling layers, and Fully Connected layers. The details of each layer are explained in this section.

\subsection{Convolution layer}

CNNs begin with a series of convolution layers, which use smaller matrices called kernels to perform convolution operations on the input image matrix~\cite{shea2015}, thereby extracting relevant spatial features. For example, consider the input image represented by the following $3\times3$ matrix:
\begin{equation}
    I=\left[\begin{array}{c c c}{{I_{00}}}&{{I_{01}}}&{{I_{02}}}\\ {{I_{10}}}&{{I_{11}}}&{{I_{12}}}\\ {{I_{20}}}&{{I_{21}}}&{{I_{22}}}\end{array}\right],
\end{equation}
and apply a kernel of size $2\times2$ with stride (the step size in horizontal or vertical position) of 1:
\begin{equation}
    K=\left[{\begin{array}{c l}{K_{00}}&{K_{01}}\\ {K_{10}}&{K_{11}}\end{array}}\right].
\end{equation}
At first, the $2\times2$ sub-matrix of $I$ at rows $0-1$, columns $0-1$ is taken, namely, $\left[\begin{array}{c c}{{I_{00}}}&{{I_{01}}}\\ {{I_{10}}}&{{I_{11}}}\end{array}\right]$, then multiply it element wise by $K$ and finally sum the four products to get $O_{00}=I_{00}K_{00}+I_{01}K_{01}+I_{10}K_{10}+I_{11}K_{11}\,$. We then slide the kernel one column to cover the sub-matrix at columns $1-2$ for $O_{01}$, one row down for $O_{10}$, and so on, until every valid position $(i, j)$ in the output image is filled. Explicitly we get, 
\begin{subequations}
    \begin{align}{{O_{00} =
I_{00}K_{00}+I_{01}K_{01}+I_{10}K_{10}+I_{11}K_{11},}}\\ {{O_{01}=I_{01}K_{00}+I_{02}K_{00}+I_{11}K_{10}+I_{12}K_{11},}}\\ {{O_{10}=I_{10}K_{00}+I_{11}K_{01}+I_{21}K_{10}+I_{21}K_{11},}}\\ {{O_{11}=I_{11}K_{00}+I_{12}K_{00}+I_{12}K_{10}+I_{22}K_{11}.}}\end{align}
\end{subequations}
Putting these together, the $2\times 2$ output image is
\begin{equation}
    O=\left[\begin{array}{c c}{{O_{00}}}&{{O_{01}}}\\ {{O_{10}}}&{{O_{11}}}\end{array}\right],
\end{equation}
where in general each entry is computed by
\begin{equation}
    O_{i j}=\sum_{u=0}^{1}\sum_{v=0}^{1}K_{u v}\,\times\,I_{i+u,\,j+v},
\end{equation}
with $(i. j)$ indexing the output map, $(u, v)$ indexing the kernel entries, and $I_{i+u,\,j+v}$ being the corresponding input pixel under $K_{u v}$. Note that for an input $I$ with dimension $H\times W$ and a kernel with size $M\times N$, the output image will have a size of $(H-M+1)\times (W-N+1)$ respectively. This whole process is illustrated in Fig.~\ref{fig:conv_operation}.
\begin{figure}[httb]
  \centering
  \includegraphics[width=0.5\linewidth]{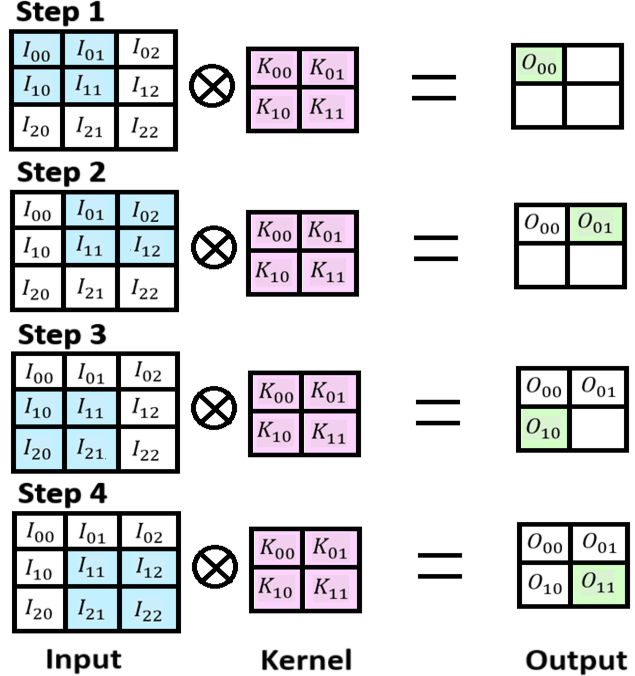}
  \caption{Convolution operation between the input image and kernels in a step-by-step manner yielding the output image.}
  \label{fig:conv_operation}
\end{figure}

\subsection{Pooling layer}

The output image which is formed after the convolution operation is called a feature map. This feature map is down-sampled by the Pooling layer using kernels of size of 2 or 3. It should be noted that using kernels of size higher than 3 may cause poor performance for the CNN. Thus it preserves the dominant features of those feature maps while shrinking their dimensionality~\cite{shea2015}. Among the various types of pooling operations such as max, min, gated, average, etc, the most common operation is the Max Pooling, which selects the highest value in each window and preserves it as the dominant feature. Figure~\ref{fig:pool_operation} shows an example of a Max Pooling operation using a $2\times2$ kernel and stride 2.

\begin{figure}[httb]
  \centering
  \includegraphics[width=0.5\linewidth]{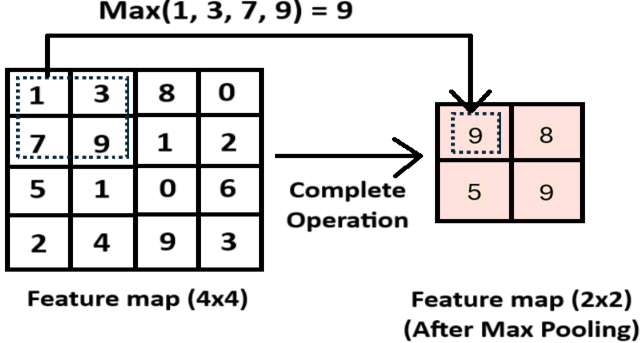}
  \caption{The dimension of the feature map is reduced from $4\times4$ to $2\times2$ during Max Pooling operation.}
  \label{fig:pool_operation}
\end{figure}

\subsection{Fully connected layer}

After the last convolutional and pooling operations, the resulting feature maps are flattened into a 1D vector. This vector is then fed into the fully connected (FC) layer for the final prediction. This layer contains several neurons, each of which is connected to the neurons in the two adjacent FC layers~\cite{alzubaidi2021}. The output is a classifier that distinguishes among various objects. In Fig.~\ref{fig:fc_layer} the architecture of the FC layer is shown.

\begin{figure}[httb]
  \centering
  \includegraphics[width=0.35\linewidth]{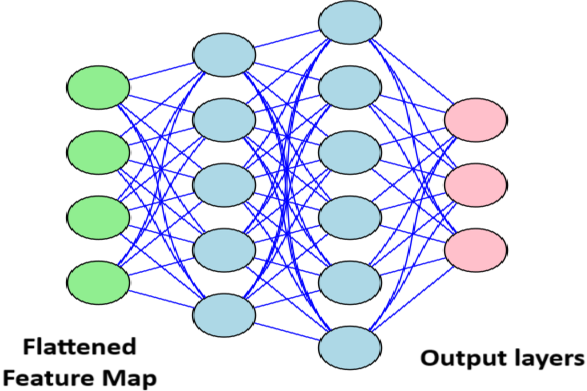}
  \caption{FC layer takes a flattened image and classifies it in its output layer respectively.}
  \label{fig:fc_layer}
\end{figure}

\newpage

\section{Appendix: Comparison table for all test samples}

\begin{longtable}{c c S S S S}
  \caption{FWHM comparison between ASTRA truth and CNN prediction for all variables across 20 test distributions.}%
  \label{tab:fwhm_longtable} \\
  \hline
  Test sample & Variable & {Truth FWHM ($f_t$)} & {Predict FWHM ($f_p$)} & {$f_t - f_p$} & {$f_p/f_t$}
  \\
  \hline
  \endfirsthead

  \hline
  Test sample & Variable & {Truth FWHM ($f_t$)} & {Predict FWHM ($f_p$)} & {$f_t - f_p$} & {$f_p/f_t$}
  \\
  \hline
  \endhead

  \hline
  \endfoot

  1  & $x$   & 4.0875   & 3.45     & 0.6375      & 0.844037 \\
  -- & $x'$  & 1875     & 1875     & 0           & 1        \\
  -- & $y$   & 2.475    & 2.8125   & -0.3375     & 1.13636  \\
  -- & $y'$  & 1937.5   & 1875     & 62.5        & 0.967742 \\
  -- & $t$   & 14.0625  & 13.725   & 0.3375      & 0.976    \\
  -- & $p_z$ & 0.745063 & 0.722312 & 0.02275     & 0.969466 \\
  2  & $x$   & 2.625    & 2.8125   & -0.1875     & 1.07143  \\
  -- & $x'$  & 1875     & 1937.5   & -62.5       & 1.03333  \\
  -- & $y$   & 3.5625   & 3.4125   & 0.15        & 0.957895 \\
  -- & $y'$  & 1875     & 1937.5   & -62.5       & 1.03333  \\
  -- & $t$   & 10.125   & 11.475   & -1.35       & 1.13333  \\
  -- & $p_z$ & 0.665438 & 0.725156 & -0.0597187  & 1.08974  \\
  3  & $x$   & 3.375    & 3.0375   & 0.3375      & 0.9      \\
  -- & $x'$  & 1875     & 1875     & 0           & 1        \\
  -- & $y$   & 2.475    & 2.6625   & -0.1875     & 1.07576  \\
  -- & $y'$  & 1937.5   & 1875     & 62.5        & 0.967742 \\
  -- & $t$   & 14.0625  & 13.5     & 0.5625      & 0.96     \\
  -- & $p_z$ & 0.745063 & 0.722312 & 0.02275     & 0.969466 \\
  4  & $x$   & 3.4875   & 3.1875   & 0.3         & 0.913978 \\
  -- & $x'$  & 1875     & 1937.5   & -62.5       & 1.03333  \\
  -- & $y$   & 2.3625   & 2.625    & -0.2625     & 1.11111  \\
  -- & $y'$  & 1875     & 1937.5   & -62.5       & 1.03333  \\
  -- & $t$   & 5.625    & 4.5      & 1.125       & 0.8      \\
  -- & $p_z$ & 0.651219 & 0.725156 & -0.0739375  & 1.11354  \\
  5  & $x$   & 3.5625   & 3.1125   & 0.45        & 0.873684 \\
  -- & $x'$  & 1875     & 2000     & -125        & 1.06667  \\
  -- & $y$   & 2.625    & 2.625    & 0           & 1        \\
  -- & $y'$  & 2187.5   & 2000     & 187.5       & 0.914286 \\
  -- & $t$   & 11.7     & 9.9      & 1.8         & 0.846154 \\
  -- & $p_z$ & 0.745063 & 0.722312 & 0.02275     & 0.969466 \\
  6  & $x$   & 3.1875   & 2.8125   & 0.375       & 0.882353 \\
  -- & $x'$  & 1875     & 1937.5   & -62.5       & 1.03333  \\
  -- & $y$   & 3.375    & 3.525    & -0.15       & 1.04444  \\
  -- & $y'$  & 1875     & 1937.5   & -62.5       & 1.03333  \\
  -- & $t$   & 10.125   & 11.25    & -1.125      & 1.11111  \\
  -- & $p_z$ & 0.665438 & 0.725156 & -0.0597187  & 1.08974  \\
  7  & $x$   & 3.075    & 2.8125   & 0.2625      & 0.914634 \\
  -- & $x'$  & 1875     & 1937.5   & -62.5       & 1.03333  \\
  -- & $y$   & 3.375    & 3.4125   & -0.0375     & 1.01111  \\
  -- & $y'$  & 1875     & 1937.5   & -62.5       & 1.03333  \\
  -- & $t$   & 12.375   & 13.5     & -1.125      & 1.09091  \\
  -- & $p_z$ & 0.7735   & 0.725156 & 0.0483438   & 0.9375   \\
  8  & $x$   & 2.85     & 2.8125   & 0.0375      & 0.986842 \\
  -- & $x'$  & 1875     & 2000     & -125        & 1.06667  \\
  -- & $y$   & 3.5625   & 4.1625   & -0.6        & 1.16842  \\
  -- & $y'$  & 1875     & 2000     & -125        & 1.06667  \\
  -- & $t$   & 6.75     & 5.4      & 1.35        & 0.8      \\
  -- & $p_z$ & 0.679656 & 0.728    & -0.0483437  & 1.07113  \\
  9  & $x$   & 3        & 2.8125   & 0.1875      & 0.9375   \\
  -- & $x'$  & 1875     & 1937.5   & -62.5       & 1.03333  \\
  -- & $y$   & 3.375    & 3.525    & -0.15       & 1.04444  \\
  -- & $y'$  & 1875     & 1937.5   & -62.5       & 1.03333  \\
  -- & $t$   & 9.1125   & 9        & 0.1125      & 0.987654 \\
  -- & $p_z$ & 0.722312 & 0.725156 & -0.00284375 & 1.00394  \\
  10 & $x$   & 2.8125   & 3.1875   & -0.375      & 1.13333  \\
  -- & $x'$  & 1875     & 2000     & -125        & 1.06667  \\
  -- & $y$   & 3.9375   & 6.6375   & -2.7        & 1.68571  \\
  -- & $y'$  & 1875     & 2062.5   & -187.5      & 1.1      \\
  -- & $t$   & 5.625    & 6.75     & -1.125      & 1.2      \\
  -- & $p_z$ & 0.759281 & 0.728    & 0.0312813   & 0.958801 \\
  11 & $x$   & 2.625    & 2.9625   & -0.3375     & 1.12857  \\
  -- & $x'$  & 1875     & 1937.5   & -62.5       & 1.03333  \\
  -- & $y$   & 4.5      & 4.2      & 0.3         & 0.933333 \\
  -- & $y'$  & 1875     & 1937.5   & -62.5       & 1.03333  \\
  -- & $t$   & 13.05    & 14.175   & -1.125      & 1.08621  \\
  -- & $p_z$ & 0.745063 & 0.725156 & 0.0199063   & 0.973282 \\
  12 & $x$   & 2.625    & 3.1875   & -0.5625     & 1.21429  \\
  -- & $x'$  & 1875     & 2125     & -250        & 1.13333  \\
  -- & $y$   & 3.9375   & 6.9      & -2.9625     & 1.75238  \\
  -- & $y'$  & 1875     & 2250     & -375        & 1.2      \\
  -- & $t$   & 6.75     & 6.75     & 0           & 1        \\
  -- & $p_z$ & 0.688188 & 0.725156 & -0.0369687  & 1.05372  \\
  13 & $x$   & 2.625    & 3        & -0.375      & 1.14286  \\
  -- & $x'$  & 1875     & 1937.5   & -62.5       & 1.03333  \\
  -- & $y$   & 3.75     & 6.0375   & -2.2875     & 1.61     \\
  -- & $y'$  & 1875     & 1937.5   & -62.5       & 1.03333  \\
  -- & $t$   & 12.9375  & 14.175   & -1.2375     & 1.09565  \\
  -- & $p_z$ & 0.716625 & 0.725156 & -0.00853125 & 1.0119   \\
  14 & $x$   & 3.9375   & 6.75     & -2.8125     & 1.71429  \\
  -- & $x'$  & 1875     & 2000     & -125        & 1.06667  \\
  -- & $y$   & 2.8125   & 3        & -0.1875     & 1.06667  \\
  -- & $y'$  & 1875     & 2000     & -125        & 1.06667  \\
  -- & $t$   & 8.4375   & 9.9      & -1.4625     & 1.17333  \\
  -- & $p_z$ & 0.665438 & 0.725156 & -0.0597187  & 1.08974  \\
  15 & $x$   & 3.75     & 6.9      & -3.15       & 1.84     \\
  -- & $x'$  & 1875     & 2000     & -125        & 1.06667  \\
  -- & $y$   & 2.625    & 3        & -0.375      & 1.14286  \\
  -- & $y'$  & 2187.5   & 2000     & 187.5       & 0.914286 \\
  -- & $t$   & 6.75     & 5.9625   & 0.7875      & 0.883333 \\
  -- & $p_z$ & 0.7735   & 0.725156 & 0.0483438   & 0.9375   \\
  16 & $x$   & 4.0875   & 6.3375   & -2.25       & 1.55046  \\
  -- & $x'$  & 1875     & 2000     & -125        & 1.06667  \\
  -- & $y$   & 2.9625   & 3        & -0.0375     & 1.01266  \\
  -- & $y'$  & 1875     & 2000     & -125        & 1.06667  \\
  -- & $t$   & 10.125   & 10.125   & 0           & 1        \\
  -- & $p_z$ & 0.745063 & 0.725156 & 0.0199063   & 0.973282 \\
  17 & $x$   & 3.75     & 6.8625   & -3.1125     & 1.83     \\
  -- & $x'$  & 1937.5   & 2187.5   & -250        & 1.12903  \\
  -- & $y$   & 2.775    & 3        & -0.225      & 1.08108  \\
  -- & $y'$  & 1875     & 2250     & -375        & 1.2      \\
  -- & $t$   & 8.4375   & 6.75     & 1.6875      & 0.8      \\
  -- & $p_z$ & 0.733688 & 0.725156 & 0.00853125  & 0.988372 \\
  18 & $x$   & 2.8125   & 3        & -0.1875     & 1.06667  \\
  -- & $x'$  & 1875     & 1937.5   & -62.5       & 1.03333  \\
  -- & $y$   & 4.6875   & 3.1875   & 1.5         & 0.68     \\
  -- & $y'$  & 1875     & 1937.5   & -62.5       & 1.03333  \\
  -- & $t$   & 8.4375   & 9        & -0.5625     & 1.06667  \\
  -- & $p_z$ & 0.787719 & 0.725156 & 0.0625625   & 0.920578 \\
  19 & $x$   & 2.8125   & 3        & -0.1875     & 1.06667  \\
  -- & $x'$  & 1875     & 1937.5   & -62.5       & 1.03333  \\
  -- & $y$   & 4.6875   & 3.225    & 1.4625      & 0.688    \\
  -- & $y'$  & 1875     & 1937.5   & -62.5       & 1.03333  \\
  -- & $t$   & 9        & 9        & 0           & 1        \\
  -- & $p_z$ & 0.747906 & 0.725156 & 0.02275     & 0.969582 \\
  20 & $x$   & 2.625    & 3        & -0.375      & 1.14286  \\
  -- & $x'$  & 1875     & 1937.5   & -62.5       & 1.03333  \\
  -- & $y$   & 5.0625   & 3.2625   & 1.8         & 0.644444 \\
  -- & $y'$  & 1875     & 1937.5   & -62.5       & 1.03333  \\
  -- & $t$   & 5.625    & 5.625    & 0           & 1        \\
  -- & $p_z$ & 0.676813 & 0.725156 & -0.0483437  & 1.07143  \\
\end{longtable}

\end{document}